\documentclass[fleqn,usenatbib]{mnras}

\usepackage{newtxtext,newtxmath}   
\usepackage[T1]{fontenc}
\usepackage{graphicx}
\usepackage{amsmath}
\usepackage{booktabs}
\usepackage{xcolor}
\usepackage{needspace}      
\usepackage{placeins}       
\makeatletter
\newenvironment{fixedtable}[2]{\par\addvspace{8pt}\noindent\begin{minipage}{\columnwidth}%
  \reset@font\small\rm\refstepcounter{table}\label{#2}%
  \SFB@maketablecaption{Table \thetable}{#1}\centering}%
  {\end{minipage}\par\addvspace{8pt}}
\newenvironment{fixedfigure}[2]{\par\addvspace{8pt}\noindent\begin{minipage}{\columnwidth}%
  \reset@font\small\rm\refstepcounter{figure}\label{#2}%
  \def\fixedfig@caption{#1}\centering}%
  {\par\SFB@makefigurecaption{Figure \thefigure}{\fixedfig@caption}\end{minipage}\par\addvspace{8pt}}
\makeatother

\providecommand{\psj}{PSJ}
\providecommand{\apjl}{ApJL}
\providecommand{\araa}{ARA\&A}

\newcommand{\figfile}[2]{\IfFileExists{#1}{\includegraphics[width=#2]{#1}}{\fbox{\parbox[c][3.5cm][c]{0.95\linewidth}{\centering\texttt{\detokenize{#1}}\par(figure file pending)}}}}

\newcommand{\Tc}{TRAPPIST-1\,c}
\newcommand{\Tb}{TRAPPIST-1\,b}
\newcommand{\Te}{TRAPPIST-1\,e}
\newcommand{\co}{CO$_2$}
\newcommand{\water}{H$_2$O}
\newcommand{\um}{\,$\mu$m}
\newcommand{\Asig}{\ensuremath{A_{\rm sig}}}
\newcommand{\dlnz}{\ensuremath{\Delta\ln\mathcal{Z}}}

\title[A calibrated contamination framework for TRAPPIST-1\,c]{Detecting atmospheres in the presence of stellar contamination: an empirically calibrated framework applied to four JWST/NIRSpec PRISM transits of TRAPPIST-1\,c}

\author[A. D. Rathcke et al.]{
Alexander~D.~Rathcke,$^{1}$\thanks{E-mail: rathcke@space.dtu.dk}
Lars~A.~Buchhave$^{1}$
\\
$^{1}$DTU Space, Technical University of Denmark, Elektrovej 328, DK-2800 Kgs.~Lyngby, Denmark\\
}

\date{Accepted XXX. Received YYY; in original form ZZZ}
\pubyear{2026}

\begin{document}
\label{firstpage}
\pagerange{\pageref{firstpage}--\pageref{lastpage}}
\maketitle

\begin{abstract}
Transmission spectroscopy of rocky exoplanets orbiting late M dwarfs is limited by stellar contamination rather than photon noise. We present a framework built for this contamination-dominated regime. It fits all transit epochs jointly, with a Gaussian process per epoch modelling the time-varying stellar contamination and a single atmospheric model shared across epochs. We use atmospheric models of a single molecule (\co, \water, or CH$_4$) reduced to its shape and scaled by one number: its peak-to-trough amplitude, the difference in transit depth between the highest and the lowest point of the model spectrum across the band, in ppm. The advantage of this approach is that it assumes nothing about which atmosphere produces a signal of that size, but instead directly measures how large a signal must be to be recovered. We calibrate this framework by injecting synthetic signals and by analysing 100 four-transit subsets of 17 archival JWST transits of the airless sibling \Tb, which quantify the spurious signals that stellar contamination alone produces. Applied to four JWST/NIRSpec PRISM transits of \Tc, our framework recovers injected \co\ signals without bias and reaches detection thresholds of 146\,ppm for moderate and 171\,ppm for strong evidence, compared with 96\,ppm for the same data if there was no stellar contamination. \co\ and CH$_4$ produce no spurious signals on the airless control planet, whereas \water\ shows a persistent stellar floor of $+55 \pm 34$\,ppm that no epoch-differential method removes. Our observations of \Tc\ are consistent with a featureless, airless baseline for all three molecules. Our 95 per cent confidence upper limit of 90\,ppm for \co-shaped signals effectively rules out clear, isothermal pure-\co\ atmospheres ($\geq$0.1\,bar) and N$_2$-dominated atmospheres with 1 per cent \co\ ($\geq$1\,bar) at the equilibrium temperature. An empirical scaling study using 2 to 10 transits of \Tb\ shows the \co\ detection threshold decreasing with the number of transits $k$ as $k^{-0.6}$, reaching 80\,ppm at ten transits. We also find that water searches become limited by the stellar floor beyond about six transits. Planetary water is therefore difficult to detect around stars like TRAPPIST-1, and any claimed detection should be treated with care because the planetary signal overlaps with water absorption in the star. \co\ atmospheres, by contrast, are detectable when we utilize our framework to account for stellar activity at the cost of roughly 1.5 to 2 times as many transits.
\end{abstract}



\section{Introduction}
\label{sec:intro}

The James Webb Space Telescope (JWST) has, for the first time, brought the atmospheres of small rocky exoplanets within reach of observational characterization. The most favourable such targets are the terrestrial planets orbiting nearby M dwarfs, where atmospheric features are amplified by an order of magnitude or more compared to similar planets around solar-type stars. Several such systems, perhaps most prominently TRAPPIST-1 \citep{Gillon2016,Gillon2017}, have become prime targets for both transmission and emission spectroscopy. Whether these planets retain atmospheres despite prolonged exposure to enhanced XUV fluxes and stellar winds is a central question for comparative planetology, for the cosmic shoreline hypothesis \citep{ZahnleCatling2017}, and ultimately for assessing the habitability of worlds around late-type hosts \citep[e.g.,][]{Shields2016}.

Initial JWST observations of terrestrial M-dwarf planets have been broadly consistent with the absence of thick atmospheres \citep[for a review, see][]{KreidbergStevenson2025}. The workhorse of these efforts has been thermal emission measured at secondary eclipse, most often with MIRI photometry in the 15\um\ band, where the planet-to-star contrast is favourable and \co\ absorbs strongly. The strength of this technique lies in diagnosing the \emph{presence} of an atmosphere. Both of its principal observable consequences, redistribution of heat to the nightside and absorption in the \co\ band, act to depress the dayside flux below that of a dark, airless surface that instantly re-radiates the absorbed starlight \citep{Koll2019,Mansfield2019,Ih2023}. Such measurements, including 15\um\ phase curves of both planets, have now been published for TRAPPIST-1\,b and c \citep{Greene2023,Ducrot2025,Zieba2023,Gillon2026} and for a growing number of other rocky planets around M dwarfs \citep{Xue2024,WeinerMansfield2024,Zhang2024,Luque2025,Wachiraphan2025,Allen2025,Fortune2025,MeierValdes2025,Holmberg2026,Xue2025,Zieba2026}, and have overwhelmingly returned dayside brightness temperatures consistent with dark, bare rock. The most intriguing exception is LHS~1478\,b, whose anomalously shallow 15\um\ eclipse is potentially suggestive of an atmosphere \citep{August2025}. Tentative indications of a volcanically replenished, SO$_2$-rich atmosphere have also been reported for the sub-Earth L~98-59\,b \citep{BelloArufe2025}. What eclipse photometry does less well is \emph{characterize}. A single broadband eclipse depth is degenerate between surface albedo, heat redistribution, and molecular absorption \citep{Mansfield2019,Ducrot2025,Hammond2025,Holmberg2026}, and so carries little information about which gases are present. Transmission spectroscopy is complementary. It probes the terminator rather than the dayside, it does not rely on the planet's energy balance, and it measures absorption as a function of wavelength, so that a feature can be attributed to a specific molecule. Its price is a small signal. On a rocky planet the feature amplitude scales with the atmospheric scale height, which is small for the high-mean-molecular-weight atmospheres expected here, and it grows only slowly with surface pressure once the band cores are opaque, so transmission constrains the presence of a thick atmosphere far better than it constrains how thick it is. It has also so far delivered fewer definitive results for these planets. The published JWST spectra are mostly featureless, and where they do show structure it is attributed to stellar contamination or is degenerate with it \citep{LustigYaeger2023,Moran2023,May2023,Lim2023}.

A central obstacle is stellar contamination. The transit light source (TLS) effect \citep{Rackham2017,Rackham2018,Rackham2019} arises because the transit depth is measured relative to the full stellar disc, while the transit chord samples only the region behind the planet. If unocculted spots, faculae, or flares have spectra that differ from that of the chord, the transmission spectrum acquires wavelength-dependent features that can mimic or mask atmospheric absorption. For cool, active M dwarfs such as TRAPPIST-1, contamination amplitudes reach several hundred to about a thousand ppm \citep{Rackham2018,Lim2023,Radica2025}, comparable to or larger than the signals expected from secondary atmospheres. Disentangling stellar from planetary signals is therefore essential, and a range of correction strategies has emerged: parametric models informed by stellar photosphere libraries \citep{Rackham2018,Wakeford2019,Garcia2022,Lim2023,Radica2025}, empirical and data-driven approaches that sidestep model spectra altogether, including out-of-transit spectral monitoring \citep{Berardo2024,RackhamDeWit2024} and the use of airless companion planets as contamination proxies \citep{TJCI2024,Rathcke2025,Allen2026}, and flexible statistical models, such as Gaussian processes (GPs), that marginalize over residual wavelength-correlated features without requiring explicit component identification \citep{Espinoza2025}. A persistent difficulty for the parametric approaches is that current model spectra of ultracool photospheres lack the fidelity required at JWST precision \citep{IyerLine2020,Wakeford2019,Garcia2022,Iyer2023,RackhamDeWit2024}, motivating the empirical and statistical routes pursued here.

TRAPPIST-1 hosts seven terrestrial planets \citep{Gillon2017,Grimm2018,Agol2021}, three within the habitable zone, making it the most important nearby laboratory for terrestrial exoplanet atmospheric studies. The host is a cool M8V dwarf \citep[M8.0 $\pm$ 0.5, $T_{\rm eff} = 2566 \pm 26$~K;][]{Gillon2016,Agol2021} with substantial photospheric heterogeneity and frequent flaring \citep{Vida2017,Morris2018,Ducrot2018,Howard2023}. JWST transmission spectroscopy of the inner planet \Tb\ revealed spectra dominated by stellar contamination \citep{Lim2023}, and for \Tc, two NIRISS/SOSS transits covering 0.6--2.85\um\ showed 100--500\,ppm contamination signatures. Marginalizing over these, \citet{Radica2025} ruled out H$_2$-dominated atmospheres and disfavoured $\gtrsim$1\,bar high-mean-molecular-weight atmospheres rich in \water, NH$_3$, or CO, leaving \co- or CH$_4$-rich compositions as the scenarios their wavelength coverage could not constrain. For \Te, \citet{Espinoza2025} demonstrated with four NIRSpec PRISM transits that current stellar modelling frameworks cannot reproduce the observed contamination features, but that marginalizing over them with GPs enables atmospheric inference, excluding cloudy, H$_2$-dominated primary atmospheres at high confidence. Secondary-atmosphere constraints are presented in the companion paper by \citet{Glidden2025}. On the emission side, the MIRI secondary eclipse of \Tc\ \citep{Zieba2023} and the 15\um\ phase curves of the two inner planets \citep{Gillon2026} are consistent with a bare rock or a thin atmosphere and disfavour thick \co-dominated envelopes, but leave composition-level constraints at the terminator to transmission spectroscopy. The 3--5\um\ region of \Tc's transmission spectrum, which contains the strong 4.3\um\ \co\ band and is exactly the regime unconstrained by the NIRISS data, has so far been observed in only a single epoch \citep{Rathcke2025}. No multi-epoch analysis, and no atmospheric constraint from this wavelength region, has been presented prior to this work.

In this paper we present a framework for detecting atmospheric features in transmission spectra dominated by stellar contamination, together with an empirical calibration of what it can and cannot do, and we apply it to four JWST/NIRSpec PRISM transits of \Tc\ from GO programme 2420 (PI: Rathcke), spanning 2022 July to 2024 July and analysed over 0.8--5.0\um. Two epochs are quiet and two are affected by stellar flares occurring near or during transit. The final epoch is the quasi-simultaneous double transit of \Tb\ and \Tc\ of \citet{Rathcke2025}, who used the airless planet b as a contamination proxy for c in that single epoch. Here we analyse the epoch independently, without using the b transit, so that all four epochs are treated alike. The questions we ask are methodological first and planetary second. Are atmospheric features, in particular \co, the dominant opacity source of plausible secondary atmospheres in this temperature regime, recoverable in the presence of TRAPPIST-1's stellar contamination? At what amplitude could a real feature be distinguished from contamination? And what do the four transits then say about \Tc?

We build on the approach that \citet{Espinoza2025} introduced for \Te, in which all epochs are fitted jointly with a wavelength-dependent GP per epoch to capture the contamination that changes from visit to visit and an atmospheric signal that is shared between all epochs. We extend it in two ways, with a null test carried out on archival PRISM transits of the airless planet \Tb\ and with an injection-recovery calibration on the real \Tc\ data, and we call the resulting approach GPTLS, a Gaussian-process treatment of the transit light source effect. The null test uses a null distribution built from repeated random four-transit subsets of 17 archival PRISM transits of \Tb\ \citep[mainly obtained by the TRAPPIST-1~e/b proxy programmes GO~6456 and GO~9256;][]{Allen2026}. Every observation of \Tb\ to date, its 15 and 12.8\um\ eclipses and its 15\um\ phase curve \citep{Greene2023,Ducrot2025,Gillon2026} as well as its transmission spectra \citep{Lim2023}, is consistent with a bare rock. Its thermal emission rules out a thick atmosphere, although a tenuous atmosphere is not excluded by the current data \citep{Gillon2026}. We adopt the working assumption that \Tb\ is airless and refer to it as such throughout the paper. Because \Tb\ orbits the same star, this control sample directly measures the distribution of spurious atmospheric signals produced by stellar contamination alone, including any component that persists across epochs, which no epoch-differential method can remove and which is otherwise degenerate with a planetary signal \citep[cf.][]{Espinoza2025}. The injection-recovery calibration then uses physically motivated atmospheric templates (pure \co, pure \water, and pure CH$_4$) to measure, on the \Tc\ data themselves, how faithfully signals of known size are recovered, how honest the quoted uncertainties are, and how large a signal must be to count as a detection. A parametric correction based on stellar model spectra, following \citet{Rackham2018} with SPHINX spectra \citep{Iyer2023} and referred to below as the stellar-model approach, provides an independent cross-check. This allows us not only to constrain the atmosphere of \Tc, but to quantify, for the first time in the TRAPPIST-1 system, the amplitude at which a real \co\ feature could be distinguished from contamination. Along the way we identify a persistent bias in \water\ recovery that affects any water-based atmospheric inference in this system. Finally, we extend the control-sample machinery into a transit-scaling study ($k=2$--10 stacked transits) that provides empirical guidance for the design of future transmission programmes.

The paper is organized as follows. Section~\ref{sec:obs} describes the observations and data reduction. Section~\ref{sec:variability} characterizes the epoch-to-epoch variability of the transmission spectra with model-independent metrics. Section~\ref{sec:framework} presents GPTLS and the two comparison models. Section~\ref{sec:null} presents the \Tb\ null test, which establishes what the framework returns when no planetary signal is present. Section~\ref{sec:injrec} checks, by injecting synthetic atmospheric signals of known size into the \Tc\ data, how faithfully the framework recovers them and how large a signal it could detect on these four epochs. Section~\ref{sec:application} applies the calibrated framework to the \Tc\ data and derives atmospheric upper limits. Section~\ref{sec:scaling} presents the transit-scaling study. Section~\ref{sec:discussion} discusses the results in the context of the broader TRAPPIST-1 atmospheric search and the implications for stellar contamination modelling in M-dwarf transmission spectroscopy, and we conclude in Section~\ref{sec:conclusions}.


\section{Observations and data reduction}
\label{sec:obs}

\subsection{JWST/NIRSpec PRISM observations of TRAPPIST-1\,c}
\label{sec:obs_t1c}

We observed four transits of \Tc\ with JWST as part of GO programme 2420 (PI: Rathcke), between 2022 July 11 and 2024 July 9. All observations used the NIRSpec instrument \citep{Jakobsen2022} in Bright Object Time Series (BOTS) mode \citep{Birkmann2022} with the PRISM/CLEAR disperser--filter combination, the S1600A1 slit, and the NRSRAPID readout pattern, providing continuous spectroscopy over the nominal 0.6--5.3\um\ PRISM bandpass at resolving power $R \sim 30$--300. Target acquisition was performed on TRAPPIST-1 itself using the wide-aperture method (WATA) with the F110W filter. The first visit used 3 groups per integration. Subsequent visits used 6 groups per integration to increase the duty cycle and the signal-to-noise ratio at the red end of the spectrum, at the cost of partial saturation of a small number of pixels near the peak of the stellar spectral energy distribution, which we handle at the reduction stage (Section~\ref{sec:frida}). Each visit spanned $\approx$4.5\,h of science time, covering the 42-min transit with $\gtrsim$1\,h of baseline on either side plus detector settling time. Table~\ref{tab:obslog} summarizes the four visits.

The final visit (2024 July 9) was deliberately scheduled to capture a quasi-simultaneous (``back-to-back'') transit of \Tb\ and \Tc, and was analysed in \citet{Rathcke2025} as a demonstration of using the airless planet b as a contamination proxy for c in a single epoch. In the present work we analyse the \Tc\ transit of that epoch independently, without using the b transit, so that all four epochs are treated homogeneously within the joint framework of Section~\ref{sec:framework}.

\begin{table*}
\centering
\caption{JWST/NIRSpec PRISM observations of \Tc\ from GO 2420. All visits used the PRISM/CLEAR disperser, the S1600A1 slit, and the NRSRAPID readout pattern. Observation 5 is the double transit with \Tb\ (Section~\ref{sec:obs_t1c}).}
\label{tab:obslog}
\footnotesize
\begin{tabular}{lccccc}
\toprule
Epoch & Date (UT) & APT Obs. & Groups/int. & $N_{\rm int}$ & Subarray \\
\midrule
1 & 2022 Jul 11 & 1 & 3 & 27\,459 & SUB512S \\
2 & 2023 Oct 29 & 3 & 6 & 15\,925 & SUB512S \\
3 & 2023 Nov 8  & 4 & 6 & 15\,925 & SUB512S \\
4 & 2024 Jul 9  & 5 & 6 & 10\,208 & SUB512  \\
\bottomrule
\end{tabular}
\end{table*}

\subsection{The TRAPPIST-1\,b control sample}
\label{sec:obs_t1b}

Our calibration strategy (Sections~\ref{sec:injrec} and \ref{sec:null}) relies on an ensemble of PRISM transits of \Tb, whose lack of a thick atmosphere is established by its thermal emission \citep{Greene2023,Ducrot2025,Gillon2026} and which we treat as airless (Section~\ref{sec:intro}) and whose transmission spectra are dominated by stellar contamination \citep{Lim2023,Rathcke2025}. We use 17 archival NIRSpec PRISM transits of \Tb, listed in Table~\ref{tab:t1b_log} (Appendix~\ref{app:t1b_log}). Fifteen were obtained by the TRAPPIST-1~e/b proxy programme of Allen \& Espinoza (GO~6456 in Cycle~3 and GO~9256 in Cycle~4, a dual-cycle proposal, with one visit observed under DD~12492), which observes close transit pairs of planets b and e in order to use b as an epoch-specific contamination proxy for e \citep{Allen2026}. These observations use an instrument configuration very similar to ours (PRISM/CLEAR, SUB512 subarray, NRSRAPID, 5 groups per integration). One further transit was captured during a PRISM observation of a TRAPPIST-1\,h transit by GO~1981 (SUB512, 5 groups per integration), and the final one is the b transit of our own 2024 July 9 double-transit epoch (Section~\ref{sec:obs_t1c}; \citealt{Rathcke2025}). All \Tb\ data were reduced and fitted with the \textsc{Frida} pipeline and light-curve modelling described below, ensuring that the control sample and the science data share systematics end-to-end.

\subsection{Data reduction with \textsc{Frida}}
\label{sec:frida}

All data were reduced with our custom-built pipeline, \textsc{Frida}
\citep{Rathcke2025,August2025}. Stage~1 uses the detector-level steps of the
official STScI \textsc{jwst} pipeline \citep{Bushouse2024} (group scaling,
data-quality initialization, saturation flagging, superbias subtraction,
linearity correction, and dark-current subtraction), supplemented by custom
group-level routines. Saturation handling is made more conservative. Any pixel
exceeding 80 per cent of the digital saturation level at a given group flags
its entire detector column at that and all subsequent groups, and any
pixel--group combination flagged in more than 20 per cent of integrations is
masked in all integrations, keeping the effective ramp length of each pixel
constant in time. \textsc{Frida} then removes pre-amplifier reset offsets at
the group level, subtracting the median of the non-illuminated detector
regions flanking the spectrum, and column-wise $1/f$ noise, estimated per
group from the unilluminated top and bottom detector rows of a
median-image-subtracted frame. No up-the-ramp jump detection is performed.
Count rates are obtained with the standard \textsc{jwst} ramp-fitting and
gain-scale steps and converted to accumulated electrons per integration.

From this point \textsc{Frida} uses the \textsc{jwst} pipeline only to assign
the wavelength solution (including the in-slit wavelength correction) and to
apply the pixel-level flat field, with the fore-optics throughput component
disabled so that the instrument response is retained. No photometric
calibration is applied and the spectra remain in detector units. Bad pixels
are identified with a custom pixel mask, supplementing the Stage~1
data-quality flags, and interpolated over together with NaN- and zero-valued
pixels. Cosmic rays are removed in the time domain. Each pixel's light curve
is compared with a Gaussian-smoothed version of itself (10-integration kernel)
and $5\sigma$ outliers are iteratively replaced with the smoothed values.
One-dimensional spectra are then obtained by optimal extraction
\citep{Horne1986}, using a spatial profile built from the median of the
out-of-transit integrations, normalized column by column, smoothed along the
dispersion direction, and with low-signal pixels nulled. The pipeline
uncertainty arrays provide the pixel variances in the extraction weights.

\subsection{Light-curve fitting and transmission spectra}
\label{sec:lcfit}

For each visit we first fitted the white light curve, obtained by summing the
extracted spectra over detector columns 125--338 ($\approx$1.58--4.37\um),
with uncertainties propagated in quadrature. This range, narrower than the wavelength range retained for the spectroscopy, was chosen because we found it to minimize the out-of-transit scatter of the white light curve. Each
spectral channel's time series was first cleaned with an iterative $5\sigma$
running-mean outlier rejection. Because the detector configuration, and
hence the integration time, differs between visits, the light curves were
binned in time using inverse-variance-weighted means of 90, 50, 50, and 30
integrations for the four epochs, respectively, yielding comparable binned
cadences, and normalized to the median of the out-of-transit data. All fits
were performed on these binned light curves, which also keeps the exact-GP
computation tractable. The transit model was computed with \textsc{batman}
\citep{Kreidberg2015}, assuming circular orbits and parameterized by the time
of mid-transit $t_0$, orbital period $P$, planet-to-star radius ratio
$R_{\rm p}/R_\star$, scaled semi-major axis $a/R_\star$, and inclination $i$,
with quadratic limb darkening sampled directly in the coefficients
$(u_1, u_2)$ under wide uniform priors on $[-3, 3]$, following the
recommendation of \citet{Coulombe2024} to avoid the transit-depth biases
introduced by physically bounded limb-darkening parametrizations. We placed Gaussian priors on $P$, $a/R_\star$,
and $i$ from \citet{Agol2021} (with symmetrized uncertainties), a Gaussian
prior of width 0.01\,d on $t_0$ centred on the predicted transit time, and a
uniform prior on $R_{\rm p}/R_\star$. For the two epochs in which a \Tb\
transit also falls within the observing window (a partial transit at the
start of the 2023 November 8 visit, caught by chance, and the deliberately
scheduled quasi-simultaneous transit of 2024 July 9, \citealt{Rathcke2025}), we
fitted a two-planet model, the product of the two \textsc{batman} light
curves, with independent orbital parameters per planet and shared limb
darkening. The systematics model comprised a linear trend in time multiplying
the transit model, with residual correlated noise captured by a Gaussian
process (\textsc{george}; \citealt{Ambikasaran2015}) with a
squared-exponential kernel, taking the product of transit model and trend as
its mean function. The log-amplitude, log-timescale, and an additional
white-noise term added in quadrature to the formal uncertainties were free
hyperparameters. Posteriors were sampled with the \textsc{dynesty} dynamic
nested sampler \citep{Speagle2020}, using multi-ellipsoid bounds and
random-slice sampling with 500 initial and batch live points, and parameters
were summarized by weighted posterior medians with 16th--84th-percentile
uncertainties. The priors are summarized in Table~\ref{tab:priors_lightcurve}.

Spectroscopic light curves were then fitted at the native pixel resolution of
the PRISM (512 spectral columns), in preference to binning in wavelength
before fitting, and prepared identically to the white
light curve (outlier rejection, time binning, and out-of-transit
normalization). The wavelength-independent parameters $t_0$, $P$,
$a/R_\star$, and $i$, and the GP amplitude and timescale, were fixed to the
white-light posterior medians, while the transit depth(s), limb-darkening
coefficients, linear-trend coefficients, and a per-channel white-noise term
were fitted in each channel with the same sampler configuration.

This yields, for each visit $v$, a native-resolution transmission spectrum
$D_v(\lambda)$ with formal uncertainties. For the analyses in this paper, we
bin the native spectra by groups of 4 pixels, giving 84 spectral bins over the 0.8--5.0\um\ analysis range with per-visit uncertainties of $\approx$70--95\,ppm bluewards and $\approx$115--195\,ppm redwards of 2\um\ (where the per-pixel flux is lowest), or $\approx$60\,ppm for the four visits combined. The four native channels are combined by
inverse-variance-weighted means, with the corresponding inverse-variance
uncertainty adopted for each bin. We restrict all
subsequent analysis to 0.8--5.0\um. The nominal PRISM bandpass extends
further on both sides, but the extreme edges are too noisy to be useful.

\subsection{Stellar activity during the observations}
\label{sec:flares}

TRAPPIST-1 flares frequently, and flares are visible in a large fraction of JWST time-series observations of the system \citep{Howard2023,Lim2023,Radica2025,Allen2026}. We inspected the white light curves and, most sensitively, the native-resolution spectroscopic channels covering H$\alpha$, in which flares stand out as sharp emission spikes. Among our four epochs, the 2023 November 8 visit shows a flare in H$\alpha$ during transit, and the 2024 July 9 visit shows activity signatures in the out-of-transit baseline \citep[see also][]{Rathcke2025}. At the light curve stage, we do not mask or try to fit these flares. Instead, our joint contamination model of Section~\ref{sec:gp_model} is left to absorb any flare-driven contamination in the transmission spectra, without epoch-specific treatment. For the 2023 November 8 epoch, Appendix~\ref{app:flare} shows the H$\alpha$ light curve and compares the transmission spectrum used here with one from a similar reduction in which the flare interval was masked.

\begin{table*}
\caption{Priors adopted in the white light-curve fits of the \Tc\ visits and
the archival \Tb\ visits, together with the marginal posterior medians and
16th--84th-percentile uncertainties for the four \Tc\ epochs.
$\mathcal{N}(\mu,\sigma)$ denotes a Gaussian prior and $\mathcal{U}(a,b)$ a
uniform prior. Gaussian priors on $P$, $a/R_\star$, and $i$ are from
\citet{Agol2021} with symmetrized uncertainties. GP and noise posteriors are
quoted as the derived amplitudes and timescales (the parameters are sampled
in the log). For $t_0$ and $P$, parentheses give the uncertainty in the last
quoted digits. Epochs 3 and 4 were fitted jointly with the accompanying \Tb\
transit (Section~\ref{sec:lcfit}). Only the \Tc\ parameters are listed here.}
\label{tab:priors_lightcurve}
\scriptsize
\setlength{\tabcolsep}{2.5pt}
\begin{tabular}{lcccccc}
\hline
Parameter & Prior (\Tc) & Prior (\Tb) & Epoch 1 & Epoch 2 & Epoch 3 & Epoch 4 \\
\hline
$t_0$ (BJD$_{\rm TDB}-2450000$) & $\mathcal{N}(t_{\rm pred}, 0.01)$ & $\mathcal{N}(t_{\rm pred}, 0.01)$ & $9772.420395(17)$ & $10247.093950(18)$ & $10256.781038(22)$ & $10501.385327(14)$ \\
$P$ (d) & $\mathcal{N}(2.421937, 0.000018)$ & $\mathcal{N}(1.510826, 0.000006)$ & $2.421937(18)$ & $2.421937(18)$ & $2.421937(18)$ & $2.421937(18)$ \\
$R_{\rm p}/R_\star$ & $\mathcal{U}(0.0344, 0.1344)$ & $\mathcal{U}(0.0359, 0.1359)$ & $0.08581^{+0.00028}_{-0.00027}$ & $0.08576^{+0.00035}_{-0.00036}$ & $0.08521^{+0.00061}_{-0.00070}$ & $0.08423^{+0.00036}_{-0.00036}$ \\
$a/R_\star$ & $\mathcal{N}(28.549, 0.171)$ & $\mathcal{N}(20.843, 0.125)$ & $28.43^{+0.09}_{-0.10}$ & $28.41^{+0.09}_{-0.11}$ & $28.55^{+0.10}_{-0.12}$ & $28.42^{+0.07}_{-0.10}$ \\
$i$ ($^\circ$) & $\mathcal{N}(89.778, 0.118)$ & $\mathcal{N}(89.728, 0.165)$ & $89.84^{+0.09}_{-0.08}$ & $89.84^{+0.09}_{-0.08}$ & $89.80^{+0.10}_{-0.08}$ & $89.84^{+0.09}_{-0.08}$ \\
$e$ & 0 (fixed) & 0 (fixed) & -- & -- & -- & -- \\
$\omega$ ($^\circ$) & 90 (fixed) & 90 (fixed) & -- & -- & -- & -- \\
\hline
\multicolumn{7}{l}{Shared within each visit} \\
\hline
$u_1$ & \multicolumn{2}{c}{$\mathcal{U}(-3, 3)$} & $0.30 \pm 0.05$ & $0.17^{+0.05}_{-0.06}$ & $0.38 \pm 0.08$ & $0.15 \pm 0.03$ \\
$u_2$ & \multicolumn{2}{c}{$\mathcal{U}(-3, 3)$} & $0.08 \pm 0.08$ & $0.30^{+0.09}_{-0.09}$ & $-0.03^{+0.11}_{-0.11}$ & $0.31^{+0.05}_{-0.05}$ \\
$c_0$ & \multicolumn{2}{c}{$\mathcal{U}(0.9, 1.1)$} & $1.00039^{+0.00017}_{-0.00035}$ & $1.000359^{+0.000054}_{-0.000055}$ & $1.000736^{+0.000094}_{-0.000081}$ & $1.000091^{+0.000099}_{-0.000095}$ \\
$c_1$ (d$^{-1}$) & \multicolumn{2}{c}{$\mathcal{U}(-0.1, 0.1)$} & $-0.0045^{+0.0024}_{-0.0015}$ & $-0.00383^{+0.00052}_{-0.00051}$ & $-0.0082^{+0.0007}_{-0.0009}$ & $-0.0005^{+0.0009}_{-0.0008}$ \\
$\sigma_{\rm GP}$ (ppm) & \multicolumn{2}{c}{$\ln\sigma_{\rm GP}$: $\mathcal{U}(-16, -5)$ [$\approx$0.1--6700\,ppm]} & $211^{+323}_{-98}$ & $92^{+24}_{-16}$ & $159^{+31}_{-22}$ & $128^{+48}_{-30}$ \\
$\tau_{\rm GP}$ (min) & \multicolumn{2}{c}{$\ln\ell$ (ln d): $\mathcal{U}(-7.5, -0.5)$ [$\approx$48\,s--14.6\,h]} & $51^{+22}_{-18}$ & $10.9^{+3.7}_{-3.4}$ & $6.1^{+2.2}_{-1.8}$ & $22.7^{+4.0}_{-3.7}$ \\
$\sigma_w$ (ppm) & \multicolumn{2}{c}{$\ln\sigma_w$: $\mathcal{U}(-15, -5)$ [$\approx$0.3--6700\,ppm]} & $10395^{+7}_{-6}$ & $9625^{+6}_{-6}$ & $10144^{+7}_{-7}$ & $8106^{+5}_{-5}$ \\
\hline
\end{tabular}
\\
\textit{Note.} The archival \Tb\ transits occur in visits that also contain a
\Te\ transit. Those visits are fitted jointly for both planets with the same
prior structure, with the \Tb\ parameters used in this work.

\end{table*}


\section{Epoch-to-epoch variability of the transmission spectra}
\label{sec:variability}

\begin{figure*}
\centering
\figfile{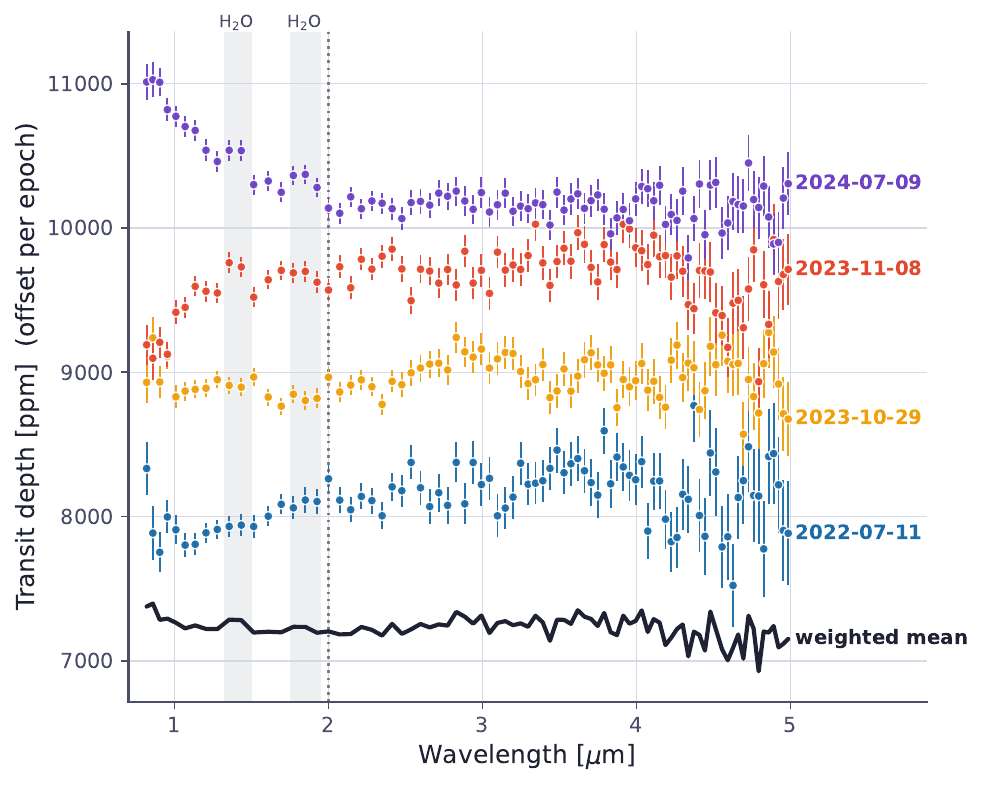}{\textwidth}
\caption{Transmission spectra of \Tc\ from the four JWST/NIRSpec PRISM epochs (4-pixel bins, 0.8--5.0\um), offset vertically by 200\,ppm per epoch for clarity and colour-coded by date, with the inverse-variance-weighted mean spectrum (black, plotted at its true depth). Grey bands mark the stellar \water\ features near 1.4 and 1.85\um. The dotted line marks 2.0\um, the boundary of the region where the contamination is strongest (Section~\ref{sec:variability}). The 2023 November 8 visit contains a flare during transit and the 2024 July 9 visit shows activity outside transit (Section~\ref{sec:flares}).}
\label{fig:spectra}
\end{figure*}

Figure~\ref{fig:spectra} presents the four transmission spectra of \Tc. Before introducing the joint modelling framework, we characterize how the spectra differ from epoch to epoch, since it is this variability, and any component that does \emph{not} vary, that determines what our contamination-correction strategy can achieve.

\subsection{Broadband variability}
\label{sec:var_broadband}

The white-light transit depths, $(R_{\rm p}/R_\star)^2$, are $7363 \pm 47$, $7355 \pm 61$, $7260 \pm 111$, and $7095 \pm 61$\,ppm for the four epochs (Table~\ref{tab:variability}), against a weighted mean of $7287 \pm 30$\,ppm, a peak-to-peak spread of 268\,ppm with $\chi^2_\nu = 4.6$ for $\nu = 3$ and an excess scatter of 110\,ppm beyond the formal uncertainties. The depths are clearly not constant from epoch to epoch. What drives the variation is less clear. An evolving heterogeneous photosphere \citep{Morris2018,Ducrot2018} is the natural candidate, but degeneracies between the depth and the freely fitted limb-darkening coefficients, the overlapping \Tb\ transit on 2024 July 9, residual systematics, or any combination of these could contribute. The broadband variation is therefore consistent with stellar contamination without being attributable to it with certainty. However, for the analysis that follows, this ambiguity does not matter, as the joint model of Section~\ref{sec:framework} fits a free depth offset $\Delta C_v$ for each visit, so the atmospheric inference uses only the shape of each spectrum and not the absolute depth.

\subsection{Chromatic variability}
\label{sec:var_metrics}
\label{sec:var_tls}

\begin{figure*}
\centering
\figfile{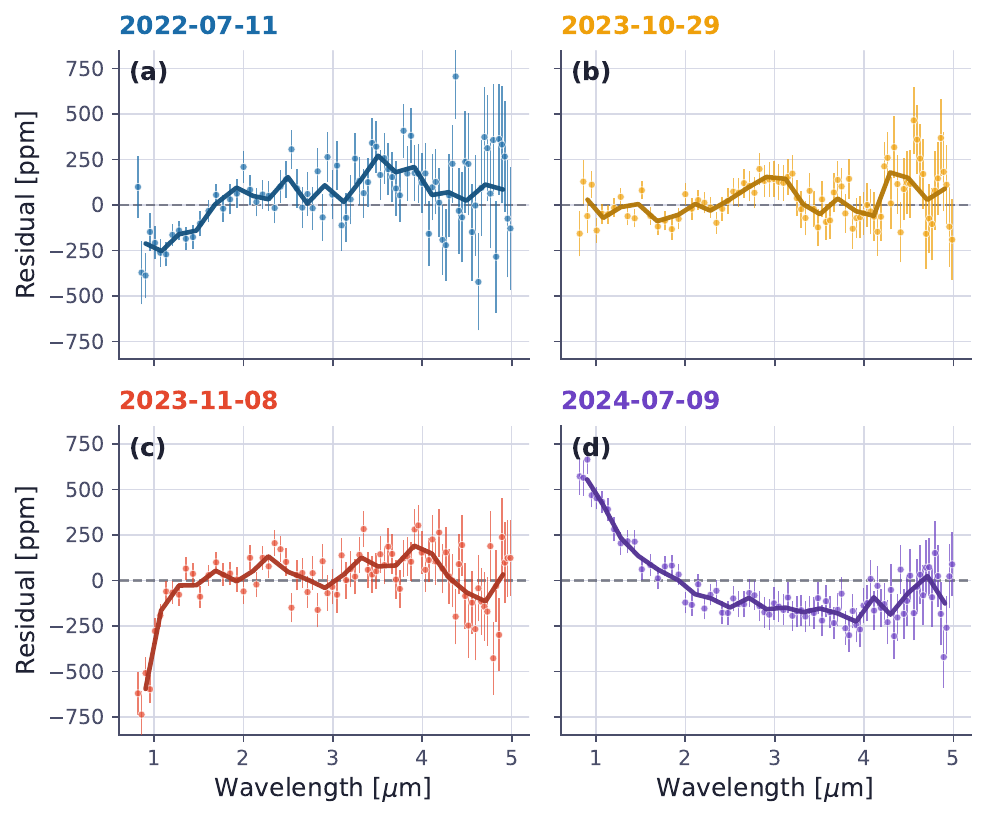}{\textwidth}
\caption{Residuals of each epoch from the weighted-mean spectrum, with one per-visit constant removed over the full 0.8--5.0\um\ range. Points are 4-pixel bins with error bars $\sqrt{\sigma^2 - 1/\Sigma w}$. The thick line is the residual re-binned to 0.2\um. Shortward of 2\um\ the residuals show smooth, epoch-dependent structure of hundreds of ppm, while longward of 2\um\ they are close to the noise expectation. Table~\ref{tab:variability} removes the constant separately within the blue and red ranges, so its $\chi^2_\nu$ values are not read directly off this figure.}
\label{fig:residuals}
\end{figure*}

The chromatic differences are what matter, and they are evident by eye in Fig.~\ref{fig:spectra}. To quantify them we proceed in two steps. First, at each wavelength bin we take the inverse-variance-weighted mean of the four epochs' depths, yielding a mean spectrum that contains everything common to all epochs (the planetary transit depth, persistent stellar features, and the instrument response). Subtracting it from each epoch, bin by bin, leaves only what differs between epochs. Second, for each epoch we average the resulting residual across the bins of a given wavelength range and subtract that single constant, so that the broadband depth offset of Section~\ref{sec:var_broadband} is discarded. This is done within the blue (0.8--2\um) and red (2--5\um) ranges separately, so that each range is tested for a common shape on its own terms. The zero-mean residuals are compared with the formal uncertainties (Fig.~\ref{fig:residuals}; Table~\ref{tab:variability}), both at the native 4-pixel binning (``fine'') and after re-binning to 0.2\um\ to isolate smooth structure (``coarse''). Bluewards of 2\um\ the four spectra are inconsistent with a common shape at the $\chi^2_\nu = 6.2$ (fine) and 18 (coarse) level, with smooth epoch-dependent structure of 140--630\,ppm peak-to-peak against a noise expectation of $\approx$150\,ppm. The two flare-affected epochs are the most discrepant, but even the nominally quiet 2022 July 11 visit carries 360\,ppm of structure. Redwards of 2\um\ the spectra are mutually consistent bin to bin ($\chi^2_\nu = 1.00$, which incidentally demonstrates that the pipeline uncertainties are well calibrated), with only mild smooth structure ($\chi^2_\nu = 1.8$ at 0.2\um\ resolution, with a peak-to-peak close to the $\approx$250\,ppm noise expectation). The red range is only weakly contaminated, though not negligibly. Our main analysis uses the full 0.8--5\um\ range, which keeps all of the band information, but as a check we also repeat it on the 2--5\um\ range alone, whose results agree with those from the full range and are summarized in Appendix~\ref{app:red}. We refer to that analysis where it is relevant, chiefly for the methane offset of Section~\ref{sec:injrec_ch4}.

\begin{table*}
\centering
\caption{Epoch-to-epoch variability metrics for the four \Tc\ transmission spectra (4-pixel binning, 0.8--5.0\um). Depth: white-light transit depth. $\chi^2_\nu$: residual of each epoch from the weighted-mean spectrum, per-visit offset removed within each range, relative to the formal uncertainties, at native binning (fine) and after re-binning to 0.2\um\ (coarse). p2p: peak-to-peak of the coarse residual. Slope: weighted linear fit over 0.8--2\um\ with the stellar water bands (1.33--1.50 and 1.80--1.95\um) masked. The joint row tests all four epochs against a common shape.}
\label{tab:variability}
\footnotesize
\begin{tabular}{lcccccc}
\toprule
 & & \multicolumn{2}{c}{Blue, 0.8--2\um} & \multicolumn{2}{c}{Red, 2--5\um} & Slope 0.8--2\um \\
\cmidrule(lr){3-4}\cmidrule(lr){5-6}
Epoch & Depth (ppm) & $\chi^2_\nu$ fine / coarse & p2p (ppm) & $\chi^2_\nu$ fine / coarse & p2p (ppm) & (ppm\,$\mu$m$^{-1}$) \\
\midrule
2022-07-11 & $7363 \pm 47$  & 3.4 / 9.1   & 360 & 0.96 / 1.31 & 261 & $+287 \pm 73$ \\
2023-10-29 & $7355 \pm 61$  & 1.7 / 1.9   & 143 & 1.01 / 2.31 & 239 & $-95 \pm 57$ \\
2023-11-08 & $7260 \pm 111$ & 8.5 / 26.3  & 626 & 1.28 / 2.17 & 308 & $+364 \pm 62$ \\
2024-07-09 & $7095 \pm 61$  & 11.2 / 34.9 & 562 & 0.74 / 1.48 & 247 & $-699 \pm 58$ \\
\midrule
Joint / mean & $7287 \pm 30$ & 6.2 / 18.0 & -- & 1.00 / 1.82 & -- & -- \\
\bottomrule
\end{tabular}
\end{table*}

The blue-end structure is dominated by epoch-dependent slopes. Weighted linear fits over 0.8--2\um\ give $+287 \pm 73$, $-95 \pm 57$, $+364 \pm 62$, and $-699 \pm 58$\,ppm\,$\mu$m$^{-1}$, a range of 1060\,ppm\,$\mu$m$^{-1}$ with the two flare-affected epochs at its extremes (the values are window-dependent because the structures are curved, and over 0.8--1.5\um\ the range is 2360\,ppm\,$\mu$m$^{-1}$). As an order-of-magnitude scale, 10 per cent of 2360\,K spots on a 2560\,K photosphere, treated as blackbodies, produce a slope of $-139$\,ppm\,$\mu$m$^{-1}$ over this window and 1 per cent of 5000\,K hot regions $+931$\,ppm\,$\mu$m$^{-1}$. The observed range thus corresponds to per-cent-level changes in hot-region coverage or to unphysically large ($\approx$80 per cent) changes in cool-spot coverage, and no single-component heterogeneity reproduces both the chromatic and the broadband variability at once. We draw no physical conclusion from this beyond the one that matters, which is that the contamination is large, epoch-dependent, and not reliably predictable from first principles. 

This is what shapes the rest of the paper. The contamination must be modelled, not ignored. Because it is smooth and structured in wavelength yet not predictable, it must be modelled flexibly and marginalized, which is what the framework of Section~\ref{sec:framework} does.


\section{Atmosphere detection frameworks}
\label{sec:framework}

Our goal is to determine whether a planetary atmospheric signal common to all epochs is present in the transmission spectra, in the presence of stellar contamination that varies from epoch to epoch and is correlated in wavelength. We cast this as a joint fit to all four transmission spectra. The planetary signal is represented by a template, a model transmission spectrum of an atmosphere made of a single molecule (\co, \water, or CH$_4$), reduced to its shape by normalizing it to unit peak-to-trough amplitude (Section~\ref{sec:templates}). The template is then scaled by a single amplitude, in ppm, that is shared by all epochs and fitted freely. This keeps the approach agnostic about how large a real feature would be, which depends on the temperature, surface pressure, mean molecular weight, and aerosol content of the atmosphere. We ask only whether a feature of that shape is present, at whatever amplitude, and translate amplitudes into atmospheric properties afterwards (Section~\ref{sec:pressure_mapping}). The contamination of each epoch is described, without any assumption about the stellar spectrum, by an independent GP in wavelength (Section~\ref{sec:gp_model}). We call this approach GPTLS, for Gaussian-process treatment of the transit light source effect. For comparison, we fit the same data with two other models. The first, which we call the offset-only model (no GP), has no GP term at all, so the contamination is absorbed only by a per-visit depth offset and a per-visit noise term (Section~\ref{sec:inference}), and the comparison isolates what the GP adds. The second is the more conventional approach of modelling the TLS effect explicitly with stellar model spectra for the photosphere and for cooler and hotter heterogeneities, with per-epoch covering fractions as free parameters, which we call the stellar-model approach (Section~\ref{sec:tls_model}). Several earlier studies have found that the model spectra do not reproduce hosts as cool as TRAPPIST-1 at the required precision \citep{Wakeford2019,Garcia2022,Lim2023,RackhamDeWit2024,Espinoza2025}, and our results confirm that.

\subsection{The Gaussian process contamination model}
\label{sec:gp_model}

For each visit $v$ we model the logarithm of the observed transit depth spectrum as
\begin{equation}
\ln D_v^{\rm obs}(\lambda) = \ln D_v^{\rm true}(\lambda) + {\rm GP}_v(\lambda) + \epsilon_v(\lambda),
\label{eq:gp_model}
\end{equation}
where the noiseless planetary spectrum is
\begin{equation}
D_v^{\rm true}(\lambda) = C_0 + \Delta C_v + \Asig\,\hat{T}(\lambda).
\label{eq:mean_model}
\end{equation}
Here $C_0$ is a shared baseline transit depth, $\Delta C_v$ is a wavelength-independent per-visit depth offset (with $\Delta C_0 \equiv 0$ for the reference visit), absorbing the broadband variability of Section~\ref{sec:var_broadband}, $\hat{T}(\lambda)$ is a normalized atmospheric template (Section~\ref{sec:templates}), and $\Asig$ is the atmospheric signal amplitude, the single parameter of interest, shared across all visits. The term ${\rm GP}_v(\lambda)$ is an independent Gaussian process in wavelength for each visit, with a Mat\'ern-3/2 covariance kernel,
\begin{equation}
k_v(\lambda_i, \lambda_j) = a_v^2 \left(1 + \frac{\sqrt{3}\,|\lambda_i - \lambda_j|}{\ell_v}\right) \exp\!\left(-\frac{\sqrt{3}\,|\lambda_i - \lambda_j|}{\ell_v}\right),
\label{eq:matern}
\end{equation}
parameterized by a per-visit amplitude $a_v$ ($=e^{\ln a_v}$) and length scale $\ell_v$, both sampled in the log. The Mat\'ern-3/2 kernel is flexible enough to describe both the smooth chromatic slopes and the broad molecular features characteristic of TLS contamination, while remaining smooth on the scale of individual spectral bins. Finally, $\epsilon_v$ denotes the observational noise, a diagonal Gaussian term combining the propagated formal per-bin uncertainties with a per-visit jitter, $s_v = e^{\ln s_v}$, added in quadrature.

The physical logic of this decomposition is that anything that varies between epochs and is correlated in wavelength is attributed to the star, while a signal that is static across epochs and matches the template shape is attributed to the planet \citep[cf.][]{Espinoza2025}. The critical caveat is that stellar contamination which persists across all epochs is indistinguishable from a planetary signal within any single system of this kind. We address it empirically with the \Tb\ control sample in Section~\ref{sec:null}.

Applied to the four \Tc\ spectra, this model yields for each visit a GP amplitude, a length scale, and a jitter (Table~\ref{tab:gp_readout}). The fitted amplitudes of 97--475\,ppm rank the epochs in the same order as the model-independent metrics of Section~\ref{sec:variability}, with 2024 July 9 and 2023 November 8 the most contaminated and 2023 October 29 the least. The length scales of 0.6--2.2\um\ match the smooth structures seen in Fig.~\ref{fig:residuals}, and the jitters of 14--24\,ppm show that, once the correlated component is modelled, the formal per-bin uncertainties underestimate the remaining uncorrelated noise by only a few per cent.

\begin{table}
\centering
\caption{Per-visit GP hyperparameters from the GPTLS fit to the four \Tc\ spectra over the full wavelength range, with no atmospheric signal component, given as posterior medians with 16th--84th-percentile ranges. Amplitudes and jitters are converted from the fitted log-depth units using $D \approx 7300$\,ppm.}
\label{tab:gp_readout}
\begin{tabular}{lccc}
\toprule
Epoch & $a_v$ (ppm) & $\ell_v$ ($\mu$m) & $s_v$ (ppm) \\
\midrule
2022-07-11 & 172 (117--276) & 1.1 (0.6--2.1) & 17 \\
2023-10-29 & 97 (65--167)   & 0.6 (0.4--1.3) & 16 \\
2023-11-08 & 273 (176--504) & 0.9 (0.5--1.6) & 24 \\
2024-07-09 & 475 (333--671) & 2.2 (1.5--2.7) & 14 \\
\bottomrule
\end{tabular}
\end{table}

For four visits, the model of equation~(\ref{eq:gp_model}) has 17 free parameters: $\Asig$, $C_0$, three visit offsets $\Delta C_v$, and per-visit jitters, GP amplitudes, and GP length scales (4 each). We also implement a reduced variant in which a single length scale $\ell$ is shared among visits (14 parameters for four visits), which we employ in the transit-scaling study of Section~\ref{sec:scaling} where the parameter count would otherwise grow impractically with the number of visits. Priors on all parameters are listed in Table~\ref{tab:priors}.

\begin{table}
\centering
\caption{Model parameters and priors for the GPTLS fit ($N_v$ visits, 17 parameters for $N_v = 4$ with per-visit length scales). $\mathcal{U}$ denotes a uniform prior. A single prior set is shared by all visits (including the flare-affected epochs), template species, and wavelength ranges.}
\label{tab:priors}
\footnotesize
\begin{tabular}{lll}
\toprule
Parameter & Description & Prior \\
\midrule
$\Asig$ (ppm) & Signal amplitude (peak-to-trough) & $\mathcal{U}(-250, 600)$ \\
$C_0$ (ppm) & Baseline transit depth & $\mathcal{U}(6500, 8500)$ \\
$\Delta C_v$ (ppm) & Per-visit depth offset & $\mathcal{U}(-500, 500)$ \\
$\ln s_v$ & Per-visit jitter (log) & $\mathcal{U}(-7, -1)$ \\
$\ln a_v$ & Per-visit GP amplitude (log) & $\mathcal{U}(-8, -1.5)$ \\
$\ln \ell_v$ & GP length scale (log; $\ell_v$ in $\mu$m) & $\mathcal{U}(-1.2, 1.1)$ \\
\bottomrule
\end{tabular}
\end{table}

\subsection{Atmospheric templates and their normalization}
\label{sec:templates}

\begin{figure}
\centering
\figfile{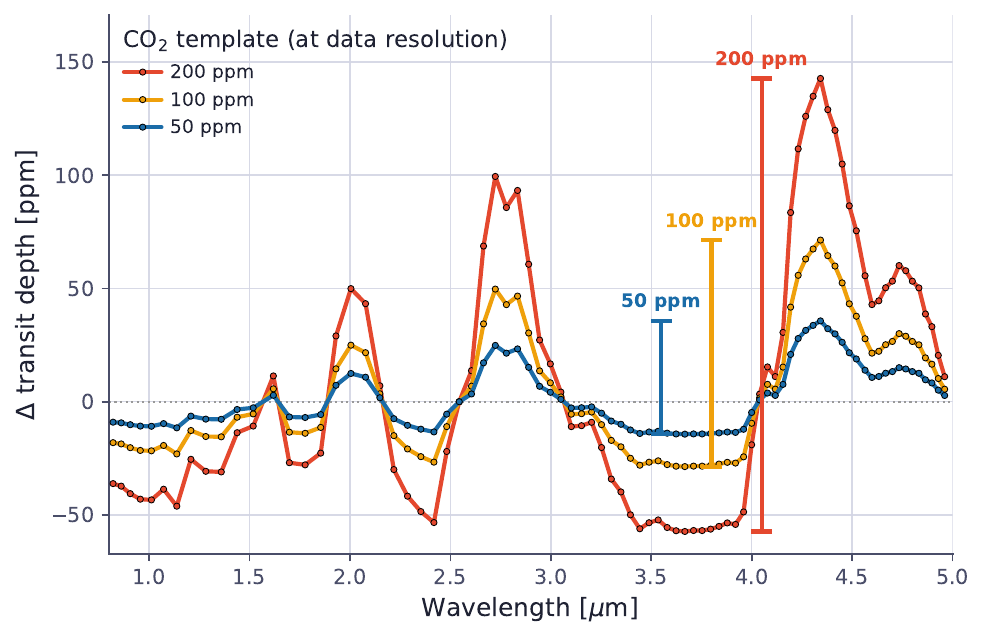}{\columnwidth}
\caption{What a peak-to-trough amplitude means. The pure-\co\ template of Section~\ref{sec:templates}, on the wavelength grid of the data, scaled to $\Asig = 50$, 100, and 200\,ppm. In each case the amplitude is the difference in transit depth between the highest and the lowest point of the model spectrum over 0.8--5\um, marked by the vertical bars.}
\label{fig:template_scale}
\end{figure}

The template $\hat{T}(\lambda)$ is generated with the open-source radiative transfer code \textsc{platon} \citep{Zhang2019,Zhang2020} for a specified atmospheric composition, assuming an isothermal temperature profile at $T=500$\,K and a reference surface pressure of 1\,bar, using the stellar parameters of \citet{Agol2021} and, since only the normalized shape enters, the mass and radius of \Tb\ for both planets. The model spectrum is computed on a fine wavelength grid, interpolated to the data bins, median-subtracted, and normalized by its peak-to-trough amplitude,
\begin{equation}
\hat{T}(\lambda) = \frac{T(\lambda) - {\rm med}\,[T]}{\max[T] - \min[T]},
\end{equation}
so that $\Asig$ measures the peak-to-trough spectral contrast of the planetary signal in ppm. This convention is used for every amplitude in the paper, so it is worth being explicit about what it means. A signal of amplitude $\Asig$ is the template shape scaled so that the difference in transit depth between the highest and the lowest point of the model spectrum, across the whole 0.8--5\um\ range at the resolution of the data, equals $\Asig$. It is not the depth of one band relative to a continuum, and not the amplitude of a sinusoid. Fig.~\ref{fig:template_scale} shows the \co\ template at 50, 100, and 200\,ppm on the wavelength grid of the data, so that the reader can see what a signal of each size looks like against the spectra of Fig.~\ref{fig:spectra}. We consider three template compositions spanning the plausible secondary-atmosphere phenomenology. Pure \co\ ($\mu = 44$) has its strongest features (2.7, 4.3\um) in the red part of the bandpass, where stellar contamination is weakest. Pure \water\ ($\mu = 18$) has bands that overlap with the water absorption present in TRAPPIST-1's own photosphere. Pure CH$_4$ ($\mu = 16$), with bands at 1.7, 2.3, and 3.3\um, together with \co\ directly targets the compositions left unconstrained by the NIRISS data \citep{Radica2025}, and its bands shortward of 2\um\ (0.9--1.7\um) fall in the region where stellar contamination is strongest. Like \co, methane has no significant absorption in the photosphere of an M8V star, whose carbon resides in CO at these temperatures, so the star is not expected to produce a persistent methane-like signal. Whether this expectation holds, both on the airless control planet and on \Tc\ itself, is tested in Sections~\ref{sec:injrec} and \ref{sec:null}.

\subsection{Bayesian inference and model comparison}
\label{sec:inference}

For each dataset we fit four models. The first two are the GPTLS model of equation~(\ref{eq:gp_model}) with the atmospheric model included (``atmosphere+GP'') and with $\Asig = 0$ (``flat+GP''). The other two are the corresponding pair without the GP term (``atmosphere'' and ``flat''), in which the contamination is absorbed only by the visit offsets and jitters, which is the offset-only model. The offset-only pair quantifies the value added by the GP marginalization. We sample all posteriors and compute Bayesian evidences with the dynamic nested sampler \textsc{dynesty} \citep{Speagle2020}, using 500 live points and a stopping criterion of $\Delta\ln\mathcal{Z}_{\rm stop} = 0.3$ for the primary \Tc\ fits (per-visit length scales), and 400 live points with $\Delta\ln\mathcal{Z}_{\rm stop} = 0.5$ for the control-sample and transit-scaling runs (shared length scale). The GP likelihood is evaluated with \textsc{celerite2} \citep{ForemanMackey2017,ForemanMackey2018}, whose $\mathcal{O}(N)$ scaling for one-dimensional inputs makes the $\sim$$10^{5}$--$10^{6}$ likelihood evaluations per fit tractable.

Our detection statistic is the Bayes factor between the atmosphere and flat variants of the same noise model,
\begin{equation}
\dlnz = \ln\mathcal{Z}_{\rm atm} - \ln\mathcal{Z}_{\rm flat},
\end{equation}
and we quote detection thresholds at $\dlnz = 3$ and $\dlnz = 5$, which lie in the ``moderate'' ($\dlnz > 2.5$) and ``strong'' ($\dlnz > 5$) bands of the Jeffreys-type scale as tabulated by \citet{Trotta2008}. Because the evidence integral penalizes the prior volume of $\Asig$, $\dlnz$ stays near zero for signals below the sensitivity of the data (Section~\ref{sec:scaling}). This built-in Occam penalty is useful, suppressing spurious low-amplitude detections.

Because $\dlnz$ carries this prior-volume penalty, we report alongside it a second, posterior-based criterion that does not, the posterior significance $z \equiv A_{\rm med}/\sigma_A$, where $A_{\rm med}$ is the posterior median of $\Asig$ and $\sigma_A = (q_{84} - q_{16})/2$ its 1$\sigma$ half-width. We quote thresholds at $z = 2$ and 3. For both criteria we also report the false-positive rate, i.e.\ how often the threshold is crossed when no planetary signal is present. This is measured directly on the airless control sample of Section~\ref{sec:null}. The evidence criteria are the stricter of the two. Because $\dlnz$ pays a penalty of $\approx$2--3 units for the prior volume of $\Asig$, a detection at $\dlnz > 5$ requires a posterior significance of $\approx$4$\sigma$ and $\dlnz > 3$ requires $\approx$3.3$\sigma$ (Section~\ref{sec:injrec_co2}). The two criteria therefore disagree for weak signals, which the posterior can resolve at 2--3$\sigma$ while the evidence remains indifferent. This matters increasingly as transits accumulate and the posterior narrows (Section~\ref{sec:scaling}).

\subsection{The stellar-model comparison approach}
\label{sec:tls_model}

As a physically motivated alternative to GPTLS, we implemented a joint-fit contamination model based on the TLS formalism of \citet{Rackham2018}, in which the contamination is built from stellar model spectra, and we refer to it as the stellar-model approach. The observed spectrum of each visit is modelled as the true planetary spectrum multiplied by a per-visit contamination factor,
\begin{equation}
D_v^{\rm obs}(\lambda) = D_v^{\rm true}(\lambda)\,
\left[1 - \sum_h f_{h,v}\left(1 - \frac{F_h(\lambda)}{F_{\rm phot}(\lambda)}\right)\right]^{-1},
\label{eq:tls}
\end{equation}
where the sum runs over unocculted heterogeneity components $h$ with covering fractions $f_{h,v}$, $F_h$ their emergent spectra, and $F_{\rm phot}$ that of the quiescent photosphere. We include two components, a cool spot component with fixed $T_{\rm spot} = 2360$\,K and $f_{{\rm spot},v} \in [0, 0.6]$, and a hot flare component approximated as a $T = 5000$\,K black body with $f_{{\rm flare},v} \in [0, 0.2]$. Photosphere and spot spectra are drawn from the SPHINX model grid \citep{Iyer2023} at the stellar parameters of \citet{Agol2021}, with $T_{\rm phot} = 2560$\,K for the photosphere, close to their $T_{\rm eff} = 2566 \pm 26$\,K. The mean model, template treatment, sampling, and model-comparison machinery are otherwise identical to Section~\ref{sec:inference}, so that GPTLS and the stellar-model approach differ only in how the contamination is described. Full details and per-visit results are given in Appendix~\ref{app:tls}.

\subsection{Injection-recovery protocol}
\label{sec:protocol}

The calibration engine used throughout Sections~\ref{sec:injrec} to \ref{sec:scaling} is injection-recovery on real data. We add a synthetic planetary signal of known size to the real spectra and see how well the framework recovers it. Given a set of observed transmission spectra, we (i) inject a synthetic signal $\Asig^{\rm inj}\,\hat{T}(\lambda)$, identical in all visits, at a known amplitude, (ii) run the full four-model fitting suite of Section~\ref{sec:inference}, (iii) record the recovered posterior on $\Asig$ and the Bayes factor $\dlnz$, and (iv) repeat over a grid of injected amplitudes. For \Tc\ the grid spans 0--500\,ppm in steps of 50\,ppm (11 amplitudes), for the \Tb\ control sample we use a reduced grid of 0, 100, and 200\,ppm, and for the transit-scaling study 0--200\,ppm in steps of 50\,ppm. Each grid is run separately for the three template compositions over the full 0.8--5.0\um\ range and, as a check, over the 2--5\um\ range (Appendix~\ref{app:red}). Because the injections are added to the \emph{real} spectra, the recovered statistics automatically incorporate the true covariance structure of TRAPPIST-1's contamination and of our reduction, rather than an assumed noise model. This is the sense in which the calibration is empirical.

From each grid (the set of injected amplitudes for one template composition on one set of spectra) we derive five summary statistics. The first asks whether the framework returns the right amplitude on average. We take the difference between the recovered and the injected amplitude at every grid point and average it over the grid, and call the result the \emph{offset}, $\beta$. Its meaning depends on the data. On a planet known to have no atmosphere, a non-zero offset means the star adds signal, and we call it a bias. On a planet that may have an atmosphere, the offset is simply the amplitude the framework returns with nothing injected, and only the comparison with an airless planet can say whether that amplitude is more likely to come from the star or from a planetary atmosphere. We therefore measure the bias on \Tb\ (Section~\ref{sec:null}) and the offset on \Tc\ (Section~\ref{sec:injrec}). The second asks whether the framework passes every injected ppm through, or instead loses a fixed fraction of whatever is injected to the contamination model. We fit a straight line to recovered versus injected amplitude and record its slope, which we call the \emph{throughput}, $\alpha$. A slope of one means nothing is absorbed, so that recovered minus injected is the same constant at every amplitude, and a slope below one means part of every injected signal is attributed to the star. The third asks whether the quoted uncertainties are honest. The \emph{coverage} is the fraction of injections for which the injected value lies inside the 68 per cent credible interval, which should be close to 68 per cent if the posterior width is right. The fourth asks how large a signal must be before it is called a detection. The \emph{detection thresholds} are the injected amplitudes at which $\dlnz$ first crosses 3 and 5, and at which $z$ first crosses 2 and 3. The fifth applies only to ensembles of subsets, where the same injected amplitude is detected in some subsets and missed in others depending on which epochs they contain. The \emph{detection probability} is the fraction of subsets in which a given amplitude yields $\dlnz > 5$, and the amplitude at which it reaches 90 per cent is the most useful sensitivity measure for planning a programme, because the crossing of the mean $\dlnz$ hides this dependence on the epochs one happens to observe.


\section{Calibration on the airless planet TRAPPIST-1\,b}
\label{sec:null}

The per-epoch GPs remove the part of the contamination that changes from visit to visit. Whatever contamination is the same in all four epochs survives, and the model has no way of telling it apart from a planetary feature (Section~\ref{sec:gp_model}). On a planet that may have an atmosphere, the amplitude the framework returns is therefore the sum of any planetary signal and this persistent stellar component, and the two cannot be separated from that planet's data alone. This is the limitation that \citet{Espinoza2025} identify for their single-planet analysis of \Te.

We address it with \Tb. It has no detectable atmosphere \citep{Greene2023,Ducrot2025,Gillon2026,Lim2023}, it orbits the same star, and 17 PRISM transits of it are available (Section~\ref{sec:obs_t1b}). We draw 100 random sets of four \Tb\ transits (matching the four of \Tc), fit each set exactly as we will fit \Tc, and record the amplitude the framework reports. Because \Tb\ has no atmosphere, every recovered amplitude is spurious, and the spread of these 100 values is the distribution of spurious signals that the star, the pipeline, and the framework together produce with four transits. We call it the null distribution, and it is the calibration that the science target cannot provide. Its mean is the bias of the framework for a given atmospheric model template, its width is the amount by which the result depends on which four epochs one happens to observe, and counting how often the airless subsets cross a detection threshold gives the false-positive rate of that criterion. We also fit every subset with the offset-only and stellar-model comparison models of Sections~\ref{sec:inference} and \ref{sec:tls_model}.

\begin{figure}
\centering
\figfile{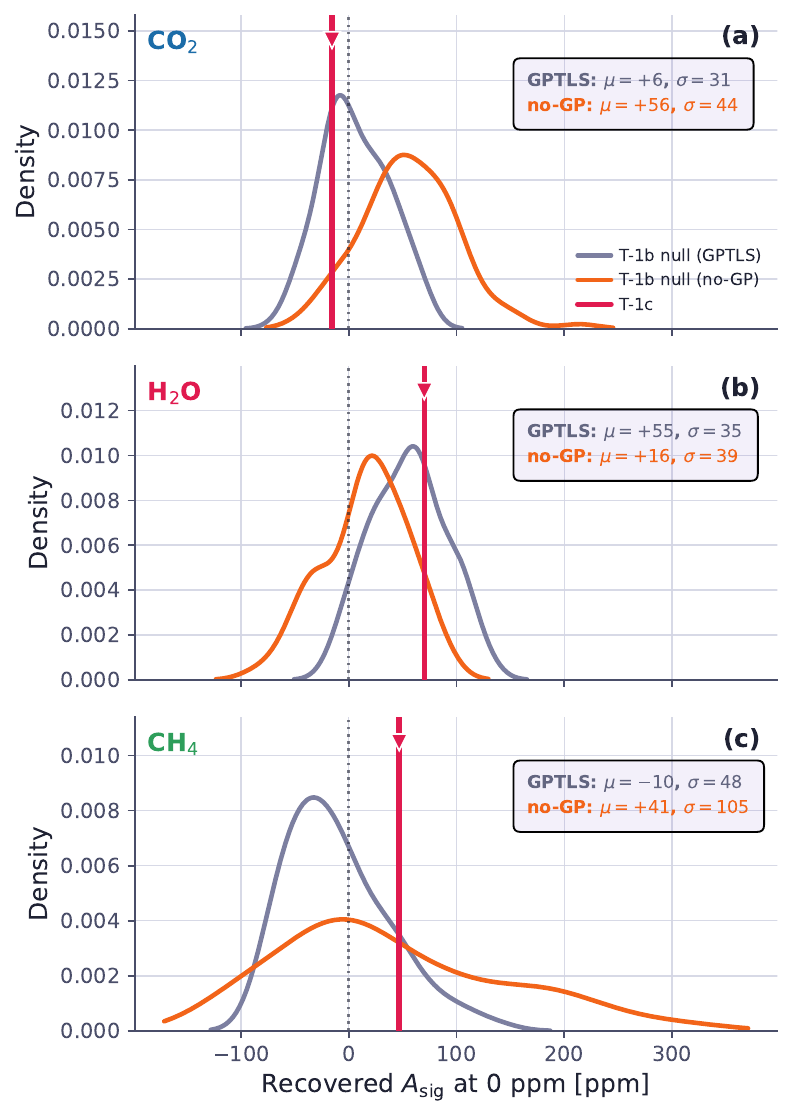}{\columnwidth}
\caption{Null distributions of the recovered amplitude at zero injection over 100 random four-transit subsets of the 17 \Tb\ transits, shown as kernel density estimates for GPTLS (grey) and the offset-only model (orange), one template per panel. The numbers in each panel are the mean and the standard deviation of the null distribution over the 100 subsets, that is, the spurious amplitude the star produces on average and how much it changes from one set of four epochs to another (Table~\ref{tab:null}). The red line and triangle mark the value recovered from the four \Tc\ transits, which lies $0.7\sigma$ below the GPTLS null mean for \co\ ($p = 0.72$), $0.4\sigma$ above for \water\ ($p = 0.33$), and $1.2\sigma$ above for CH$_4$ ($p = 0.13$). The displaced centre of the \water\ panel is the stellar water floor. The displaced offset-only curve in the \co\ panel is the spurious \co\ signal that the model without a GP produces on an airless planet (Section~\ref{sec:null}).}
\label{fig:null}
\end{figure}

\begin{table*}
\centering
\caption{The \Tc\ amplitudes compared with the airless null, for GPTLS (bold) and the two comparison models. $A_{\rm c}$ is the amplitude recovered on the four \Tc\ transits with nothing injected. $\mu_{\rm b}$ and $\sigma_{\rm b}$ are the mean and standard deviation of the amplitudes recovered, with the same model, on 100 random four-transit subsets of the airless \Tb: the null distribution. The offset of \Tc\ from the null is given in units of the null scatter, $(A_{\rm c} - \mu_{\rm b})/\sigma_{\rm b}$, and $p$ is the fraction of the 100 airless subsets on which the recovered amplitude was at least as large as \Tc's. The last three columns are the fraction of airless subsets that would have been called a detection on the posterior criterion ($z > 2$, $z > 3$) and on the evidence criterion ($\dlnz > 3$), i.e.\ the false-positive rates. For the evidence criterion only subsets with a positive recovered amplitude are counted, since a negative absorption feature is unphysical; for GPTLS this changes nothing, whereas the stellar-model approach, whose \co\ and \water\ null distributions are displaced to negative amplitudes, would reach $\dlnz > 3$ in 14 and 24 per cent of subsets if either sign were counted. All rows use the full 0.8--5\um\ range except those marked 2--5\um, which repeat the analysis on the 2--5\um\ range alone (Appendix~\ref{app:red}); all amplitudes are peak-to-trough (Section~\ref{sec:templates}). Each $\mu_{\rm b}$ carries an additional $\approx$15\,ppm uncertainty from the finite 17-epoch pool (Appendix~\ref{app:validation}).}
\label{tab:null}
\scriptsize
\setlength{\tabcolsep}{3pt}
\begin{tabular}{llrrrrrccc}
\toprule
Atmospheric & Contamination & \Tc & \multicolumn{2}{c}{Airless \Tb\ (100 subsets)} & \multicolumn{2}{c}{\Tc\ against null} & \multicolumn{3}{c}{False-positive rate on \Tb} \\
\cmidrule(lr){3-3}\cmidrule(lr){4-5}\cmidrule(lr){6-7}\cmidrule(lr){8-10}
model & model & $A_{\rm c}$ [ppm] & $\mu_{\rm b}$ [ppm] & $\sigma_{\rm b}$ [ppm] & Offset [$\sigma_{\rm b}$] & $p$ & $z>2$ & $z>3$ & $\dlnz>3$ \\
\midrule
\bfseries \co\      & \bfseries GPTLS & \bfseries $-15.5$  & \bfseries $+6.1$  & \bfseries 31.2  & \bfseries $-0.69$ & \bfseries 0.72 & \bfseries 0.00 & \bfseries 0.00 & \bfseries 0.00 \\
\co\                & offset-only     & $-9.0$   & $+56.2$ & 43.9  & $-1.48$ & 0.93 & 0.33 & 0.07 & 0.06 \\
\co\                & stellar-model   & $-196.2$ & $-53.0$ & 34.9  & $-4.10$ & 1.00 & 0.00 & 0.00 & 0.00 \\
\bfseries \water\   & \bfseries GPTLS & \bfseries $+70.1$  & \bfseries $+55.4$ & \bfseries 34.4  & \bfseries $+0.43$ & \bfseries 0.33 & \bfseries 0.32 & \bfseries 0.04 & \bfseries 0.02 \\
\water\             & offset-only     & $+30.6$  & $+15.8$ & 39.1  & $+0.38$ & 0.36 & 0.06 & 0.00 & 0.00 \\
\water\             & stellar-model   & $-169.7$ & $-69.0$ & 49.6  & $-2.03$ & 0.98 & 0.00 & 0.00 & 0.00 \\
\bfseries CH$_4$    & \bfseries GPTLS & \bfseries $+46.9$  & \bfseries $-9.5$  & \bfseries 47.4  & \bfseries $+1.19$ & \bfseries 0.13 & \bfseries 0.01 & \bfseries 0.00 & \bfseries 0.00 \\
CH$_4$              & offset-only     & $+90.2$  & $+40.7$ & 104.3 & $+0.47$ & 0.28 & 0.29 & 0.20 & 0.19 \\
CH$_4$              & stellar-model   & $+63.0$  & $+24.6$ & 48.7  & $+0.79$ & 0.25 & 0.16 & 0.02 & 0.01 \\
\bfseries CH$_4$, 2--5\um\ & \bfseries GPTLS & \bfseries $+11.7$  & \bfseries $+3.8$  & \bfseries 57.3  & \bfseries $+0.14$ & \bfseries 0.40 & \bfseries 0.05 & \bfseries 0.02 & \bfseries 0.01 \\
CH$_4$, 2--5\um\    & offset-only     & $+57.6$  & $-0.5$  & 49.8  & $+1.17$ & 0.14 & 0.08 & 0.00 & 0.00 \\
CH$_4$, 2--5\um\    & stellar-model   & $+116.0$ & $+99.1$ & 59.1  & $+0.29$ & 0.40 & 0.59 & 0.32 & 0.28 \\
\bottomrule
\end{tabular}
\end{table*}

Table~\ref{tab:null} and Fig.~\ref{fig:null} show the result, and four properties matter.

First, for \co\ and methane the null distributions are centred on zero. The star produces no persistent \co- or methane-like signal, so for these molecules the framework is unbiased, and an amplitude recovered on another planet of the same star can be taken at face value. (The centre of each distribution is itself uncertain by about 15\,ppm, because the 100 subsets are drawn from only 17 epochs, see Appendix~\ref{app:validation}. The \co\ and methane centres are consistent with zero at that level, whereas the water centre below is not.)

Second, for water the null distribution is not centred on zero. The airless \Tb\ yields a spurious water amplitude of $+55 \pm 34$\,ppm. This is the transit light source effect at work. Both the photosphere of an M8V star and its cooler spots have deep water bands, but of different strength, so whenever the unocculted disc differs in spot coverage from the transit chord, the ratio of their spectra imprints water bands onto the transit depth. Because such heterogeneity is present in every epoch, some of it survives the per-epoch GPs and projects onto the water template. We call this the stellar water floor. It is a distribution rather than a fixed number. Its value depends on which epochs one happens to observe, so any water amplitude measured on another planet of this star is a single draw from the sum of the floor and whatever the planet contributes, and cannot be separated into the two parts. The width of the floor distribution is thus the irreducible systematic for any water measurement, and it is why 32 per cent of the airless subsets show a spurious water signal above 2$\sigma$ in the posterior (Table~\ref{tab:null}).

Third, the widths of the null distributions, 30 to 50\,ppm, are a direct measurement of how much the recovered amplitude changes from one set of four epochs to another. Once a small correction for the overlap between subsets is applied (two random sets of four out of 17 epochs share, on average, about a quarter of their epochs, which makes the spread look slightly smaller than it would be for independent sets, see Appendix~\ref{app:scaling_details}), this spread agrees with the posterior width the fits themselves report to within about 20 per cent (36 against 41\,ppm for \co). The framework's error bars are therefore honest, and if anything slightly conservative. On data without contamination the posterior width is 25\,ppm (Appendix~\ref{app:validation}), so the widening on the real data is the price of allowing for GP realizations that could mimic the template. It is paid in the width of the uncertainty, not in a bias.

Fourth, the null test shows how the comparison models behave when there is no atmospheric signal to find. The offset-only model returns a spurious \co\ amplitude of $+56 \pm 44$\,ppm on the airless \Tb, and its methane null is unusable, with a width of more than 100\,ppm. On the same airless subsets it exceeds $z = 2$ in 33 per cent of cases for \co\ and 29 per cent for methane, and $\dlnz = 3$ in 6 and 19 per cent, against 0 to 1 per cent for GPTLS (Table~\ref{tab:null}). The stellar-model approach returns null distributions displaced by 25 to 70\,ppm, in a direction that depends on the template (Table~\ref{tab:null}; Appendix~\ref{app:tls}). What these failures mean for calibration on the science target is discussed in Section~\ref{sec:injrec_comparison}.

The 17 PRISM \Tb\ archive can also be used to measure detection thresholds, and Section~\ref{sec:scaling} does so with the same approach. Random sets of transits are drawn from the 17 archival ones, now with an increasing number of transits per set, synthetic signals are injected into each set, and the amplitude at which they become detectable is recorded. For sets of four transits, matching the \Tc\ dataset, a \co\ feature needs 112\,ppm to reach $\dlnz > 3$ and 135\,ppm to reach $\dlnz > 5$. These are the thresholds a typical set of four epochs of this star affords, and the reference against which the four \Tc\ epochs are judged next.


\section{Response to injected signals on TRAPPIST-1\,c}
\label{sec:injrec}

\begin{figure*}
\centering
\figfile{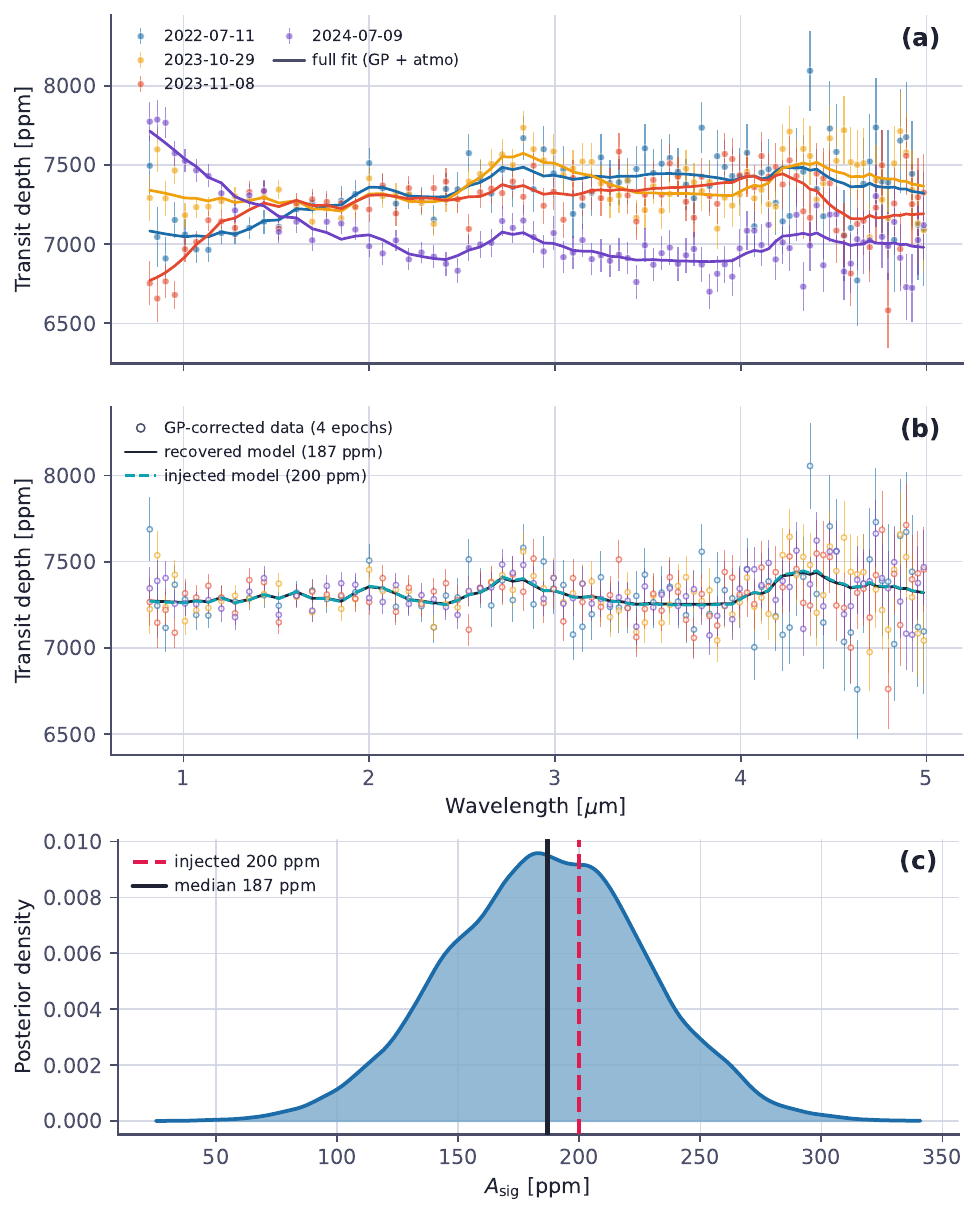}{0.72\textwidth}
\caption{Anatomy of one injection-recovery experiment. The same 200\,ppm pure \co\ template is injected into all four \Tc\ epochs and fitted jointly, with a shared amplitude and baseline and a GP and depth offset per visit. (a) The four injected spectra, colour-coded by epoch as in Fig.~\ref{fig:spectra}, each with its full fitted model (thin solid line). (b) The same four spectra after subtracting each epoch's own GP and depth offset, so that they collapse onto the shared model; the recovered atmospheric model (black) and the injected one (teal dashed) nearly coincide, because the recovery is unbiased for \co. (c) The posterior on \Asig, with the injected value and the posterior median ($\approx$187\,ppm) marked.}
\label{fig:anatomy}
\end{figure*}

\begin{figure}
\centering
\figfile{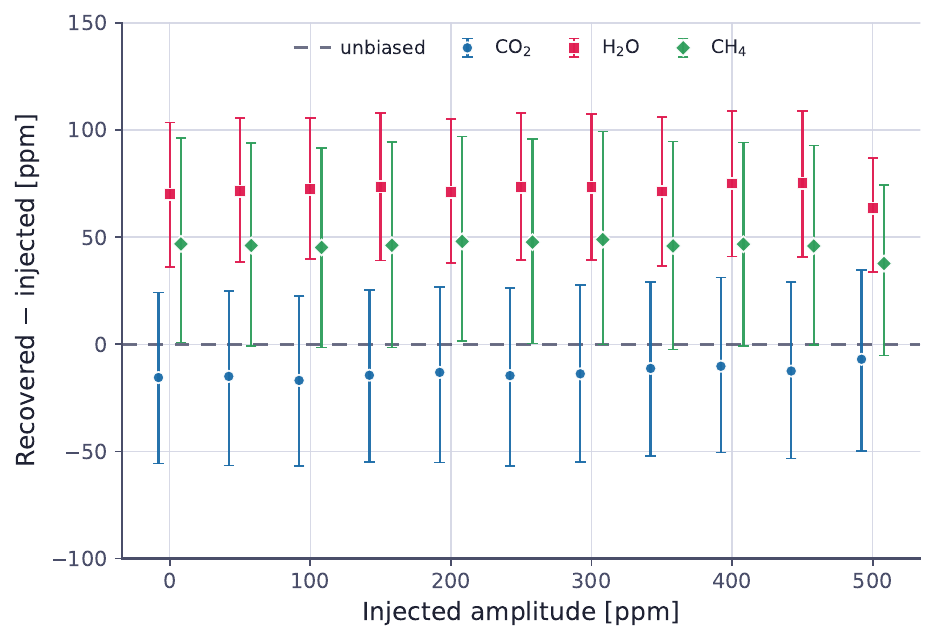}{\columnwidth}
\caption{Recovery bias for pure \co, \water, and CH$_4$ injections into the four \Tc\ epochs (GPTLS), shown as recovered minus injected amplitude versus injected amplitude, with the three templates offset slightly in the horizontal direction for legibility. Error bars span the 16th--84th posterior percentiles. The dashed line marks an unbiased recovery. \co\ is recovered without bias, whereas \water\ and CH$_4$ sit at a constant positive offset (Sections~\ref{sec:injrec_h2o} and \ref{sec:injrec_ch4}; Table~\ref{tab:injrec}).}
\label{fig:injrec}
\end{figure}

\begin{figure}
\centering
\figfile{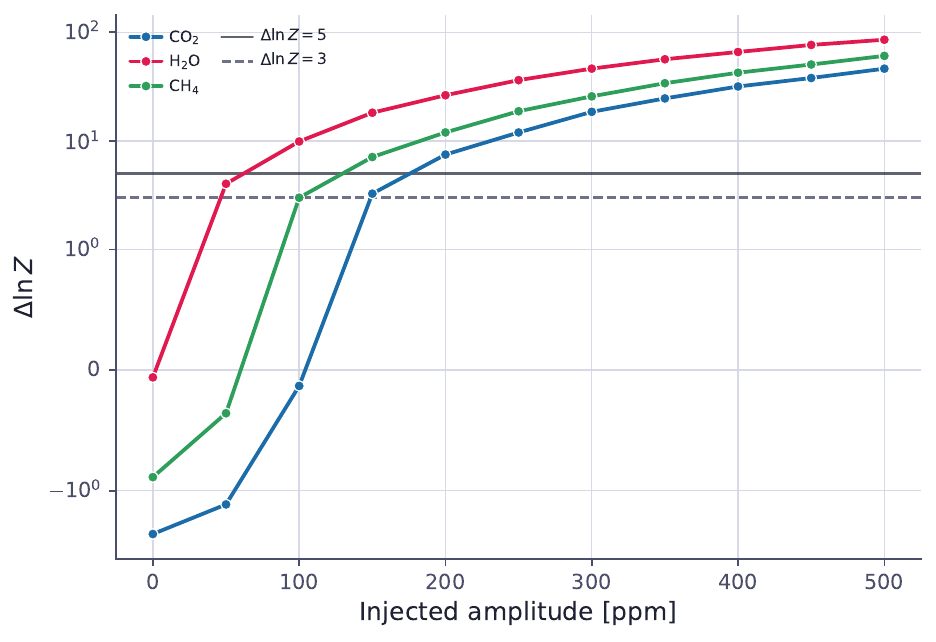}{\columnwidth}
\caption{Detection significance $\dlnz$ versus injected amplitude for the three templates (GPTLS, four \Tc\ epochs). The vertical axis is linear below $\dlnz = 1$ and logarithmic above. Horizontal lines mark $\dlnz = 3$ and 5. The \water\ and CH$_4$ crossings are offset-assisted (Sections~\ref{sec:injrec_h2o} and \ref{sec:injrec_ch4}).}
\label{fig:dlnz}
\end{figure}

Section~\ref{sec:null} established what the framework returns when there is no atmosphere. Before comparing \Tc\ with that null, we check on the four \Tc\ epochs themselves that the framework responds to a real signal as it should. We inject each of the three templates of Section~\ref{sec:templates} (pure \co, pure \water, and pure CH$_4$) at amplitudes from 0 to 500\,ppm, fit each injected data set with GPTLS and, for comparison, with the offset-only and stellar-model approaches, and measure the throughput, the coverage, and the detection thresholds defined in Section~\ref{sec:protocol}. The same grids restricted to 2--5\um\ are summarized in Appendix~\ref{app:red}. The offset of each grid, the amplitude the framework returns on \Tc\ with nothing injected, is stated here but interpreted only in Section~\ref{sec:application}, where it is compared with the null distribution. Figure~\ref{fig:anatomy} illustrates a single injection-recovery experiment, and Figs~\ref{fig:injrec} and \ref{fig:dlnz} and Table~\ref{tab:injrec} summarize all the grids.

\begin{table*}
\centering
\caption{Injection-recovery performance of GPTLS (bold) and of the two comparison models on the four \Tc\ transits. All rows use the full 0.8--5\um\ range except those marked 2--5\um, which repeat the analysis on the 2--5\um\ range alone (Appendix~\ref{app:red}); all amplitudes are peak-to-trough (Section~\ref{sec:templates}). Offset $\beta$: mean of recovered minus injected amplitude across the injection grid, which for unit throughput equals the amplitude returned with nothing injected. Throughput $\alpha$: slope of recovered against injected amplitude. $\sigma_A$: mean 1$\sigma$ posterior half-width. Coverage: fraction of injections with the injected value inside the 68 per cent credible interval. Thresholds: injected amplitude at which the Bayes factor crosses $\dlnz = 3$ and 5, and at which the posterior significance crosses $z = 3$. FP: number of zero-injection runs with $\dlnz > 3$. $^\dagger$Offset-assisted: a positive offset pushes the criterion over the line early, so these entries are not sensitivities (Section~\ref{sec:injrec_h2o}); $0^\dagger$: the criterion is already exceeded with nothing injected. ``--'': no crossing within the grid. The last three rows are the same experiment on white-noise spectra with the real error bars but no contamination (Appendix~\ref{app:validation}).}
\label{tab:injrec}
\scriptsize
\setlength{\tabcolsep}{3pt}
\begin{tabular}{llrrrrrrc}
\toprule
Atmospheric & Contamination & \multicolumn{4}{c}{Recovery} & \multicolumn{2}{c}{Detection threshold [ppm]} & \\
\cmidrule(lr){3-6}\cmidrule(lr){7-8}
model & model & $\beta$ [ppm] & $\alpha$ & $\sigma_A$ [ppm] & Coverage & $\dlnz>3$ / $>5$ & $z>3$ & FP \\
\midrule
\bfseries \co\      & \bfseries GPTLS & \bfseries $-13$  & \bfseries 1.01 & \bfseries 41 & \bfseries 100\% & \bfseries 146 / 171 & \bfseries 135 & \bfseries 0 \\
\co\                & offset-only     & $-8$   & 1.00 & 39 & 100\% & 134 / 157 & 124 & 0 \\
\co\                & stellar-model   & $-204$ & 0.97 & 25 & 0\%   & 282 / 299 & 281 & 1 \\
\bfseries \water\   & \bfseries GPTLS & \bfseries $+72$  & \bfseries 1.00 & \bfseries 33 & \bfseries 0\%   & \bfseries 37 / 58$^\dagger$ & \bfseries 30$^\dagger$ & \bfseries 0 \\
\water\             & offset-only     & $+30$  & 1.00 & 40 & 100\% & 106 / 129 & 92  & 0 \\
\water\             & stellar-model   & $-178$ & 0.96 & 30 & 0\%   & 272 / 291 & 269 & 1 \\
\bfseries CH$_4$    & \bfseries GPTLS & \bfseries $+46$  & \bfseries 0.99 & \bfseries 47 & \bfseries 73\%  & \bfseries 100 / 124$^\dagger$ & \bfseries 95$^\dagger$ & \bfseries 0 \\
CH$_4$              & offset-only     & $+88$  & 0.97 & 48 & 0\%   & 64 / 92$^\dagger$ & 58$^\dagger$ & 0 \\
CH$_4$              & stellar-model   & $+57$  & 0.97 & 28 & 0\%   & 30 / 51$^\dagger$ & 22$^\dagger$ & 0 \\
\bfseries CH$_4$, 2--5\um\ & \bfseries GPTLS & \bfseries $+14$  & \bfseries 1.00 & \bfseries 45 & \bfseries 100\% & \bfseries 131 / 156 & \bfseries 116 & \bfseries 0 \\
CH$_4$, 2--5\um\    & offset-only     & $+57$  & 0.99 & 34 & 0\%   & 56 / 76$^\dagger$ & 49$^\dagger$ & 0 \\
CH$_4$, 2--5\um\    & stellar-model   & $+114$ & 0.97 & 31 & 0\%   & $0^\dagger$ / 6$^\dagger$ & $0^\dagger$ & 1 \\
\midrule
\multicolumn{9}{l}{White-noise spectra, same error bars, no contamination (Appendix~\ref{app:validation})} \\
\co\      & GPTLS & $-7$   & 1.00 & 25 & 100\% & 79 / 96   & 79 & 0 \\
\water\   & GPTLS & $+9$   & 1.00 & 25 & 100\% & 74 / 94   & 65 & 0 \\
CH$_4$    & GPTLS & $+8$   & 1.00 & 28 & 100\% & 84 / 102  & 77 & 0 \\
\bottomrule
\end{tabular}
\end{table*}

\subsection{Carbon dioxide}
\label{sec:injrec_co2}

For the pure \co\ template the framework behaves as one would hope. The recovered amplitude rises one-for-one with the injected amplitude (Fig.~\ref{fig:injrec}), so the GP does not absorb any of the planetary signal, and the quoted uncertainties are honest, if anything slightly conservative, since the injected value lies inside the 68 per cent credible interval in every experiment. With these four transits, a \co\ feature needs a peak-to-trough amplitude of about 150\,ppm to reach moderate evidence ($\dlnz > 3$) and about 170\,ppm to reach strong evidence ($\dlnz > 5$). On the posterior criterion a 3$\sigma$ detection requires 135\,ppm, so the evidence criteria are stricter, corresponding to roughly 3.3$\sigma$ and 4$\sigma$, as expected from the Occam penalty of Section~\ref{sec:inference}. These thresholds are higher than the 112 and 135\,ppm of a typical four-epoch set from the \Tb\ archive (Section~\ref{sec:null}), likely because two of the \Tc\ epochs are flare-affected. We can put a number on what the contamination costs. For each of the four epochs, we build a synthetic transmission spectrum that keeps the observed wavelength grid and error bars but replaces the depths with a constant plus Gaussian noise drawn from those error bars, so that it contains white noise and nothing else, and then repeat the injections on these synthetic spectra (Appendix~\ref{app:validation}). On them, the strong-detection threshold is 96\,ppm and the posterior width 25\,ppm, against 171 and 41\,ppm on the real spectra. The contamination therefore costs these particular epochs a factor of 1.8 in sensitivity, against about 1.2 for typical epochs of the archive. The offset of the \co\ grid is $-13$\,ppm, close to the null mean of $+6$\,ppm measured on \Tb\ and small compared with the posterior width. Its interpretation follows in Section~\ref{sec:app_null}.

\subsection{Water}
\label{sec:injrec_h2o}

For water the response is also one-for-one, but the whole line is shifted. The offset is $+72$\,ppm, so whatever amplitude we inject, the fit returns about 72\,ppm more. This is the stellar water floor of Section~\ref{sec:null} seen on \Tc. The airless \Tb\ returns $+55 \pm 34$\,ppm, and \Tc\ returns $+70$\,ppm with nothing injected, a value consistent with a draw from that distribution. Because the floor is about twice the posterior width, the credible intervals never contain the injected value (a coverage of 0 per cent). The intervals are honest about the planetary amplitude plus the floor, not about the planetary amplitude alone.

The floor also makes the water detection thresholds misleading. Any injected water signal is added on top of the stellar water that is already in the spectra, so the fit responds to the sum of the two. The evidence crosses $\dlnz = 3$ for an injected amplitude of only 37\,ppm, and the fit to the real \Tc\ data with nothing injected already registers at 2$\sigma$ in the posterior. The stellar water thus lowers the thresholds, but for the wrong reason, since most of the signal that crosses them comes from the star, not from the planet. These thresholds, marked with a dagger in Table~\ref{tab:injrec}, are therefore not a measure of the sensitivity to planetary water. That sensitivity is best judged from the same data without contamination (Appendix~\ref{app:validation}), which gives a strong-detection threshold of 94\,ppm. The practical consequence is that a claimed water detection on a TRAPPIST-1 planet from transmission spectroscopy carries a systematic uncertainty of several tens of ppm unless the stellar floor is measured independently.

\subsection{Methane}
\label{sec:injrec_ch4}

Methane ought to be an easy molecule. It has bands across the whole bandpass (near 0.9, 1.15, 1.4, 1.7, 2.3, and 3.3\um), the star has none, and the airless planet accordingly shows no methane floor (Section~\ref{sec:null}). Yet on \Tc\ the response, again one-for-one, is shifted by a constant $+46$\,ppm, about one posterior width, and the coverage drops to 73 per cent (Table~\ref{tab:injrec}). Since \Tb\ shows nothing of the kind, this offset is not a property of the star but of these four epochs. The reason is where methane's bands lie. Carbon dioxide's strongest bands sit at 2.7 and 4.3\um, in the weakly contaminated part of the spectrum, whereas several of methane's bands, spanning many spectral bins, sit shortward of 2\um, where the contamination is strongest and changes most from epoch to epoch (Section~\ref{sec:variability}). Whatever epoch-specific structure the four \Tc\ epochs happen to share there projects onto the template as if it were a signal. Two checks support this reading. Restricting the fit to 2--5\um, which removes the blue bands but keeps the 2.3 and 3.3\um\ ones, removes the offset entirely (offset $+14$\,ppm and full coverage, Appendix~\ref{app:red}), and on data without contamination methane is recovered as cleanly as \co\ (Appendix~\ref{app:validation}). Whether the offset could nonetheless be planetary is assessed against the null distribution in Section~\ref{sec:application}. The methane thresholds (100 and 124\,ppm) are correspondingly marked as offset-assisted in Table~\ref{tab:injrec}, and the sensitivity to a planetary signal is best judged from the contamination-free data, 102\,ppm at $\dlnz > 5$.

\subsection{The comparison models}
\label{sec:injrec_comparison}

The two comparison models teach different lessons. The offset-only model, in which the contamination is absorbed only by the per-visit offsets and jitters, fails visibly for methane, where its offset is almost 90\,ppm and its credible intervals never contain the injected value, and its modest offset for water ($+30$\,ppm) is an accidental cancellation between absorbed signal and the stellar floor (Table~\ref{tab:injrec}). For \co, however, it passes every test on \Tc. Its response is one-for-one, its offset is $-8$\,ppm, its coverage is complete, and its thresholds are even slightly lower than those of GPTLS. Judged from \Tc\ alone, one would conclude that the GP is unnecessary for \co. Section~\ref{sec:null} has already shown that this conclusion is wrong. On the airless planet, the same model returns a \co\ signal of $+56 \pm 44$\,ppm that is not there. This is the central methodological lesson of the paper. Injecting a signal tests how a model responds to an added feature, not whether the model already contains a false one, so a persistent floor is invisible to any calibration performed on the science target. Only a planet known to be airless can reveal it. Conversely, on data without contamination GPTLS and the offset-only model perform identically (Appendix~\ref{app:validation}), i.e., the GP costs nothing when there is nothing for it to do.

The stellar-model approach fails for all three templates. Its offsets are $-204$\,ppm for \co, $-178$\,ppm for water, and $+57$\,ppm for methane, a sign that depends on the template. It absorbs up to 4 per cent of the injected signal, its credible intervals never contain the injected value, and it reports spurious detections with nothing injected for \co\ and water (Table~\ref{tab:injrec}). With only two fixed-temperature components, the model can reproduce the observed contamination only by pushing the covering fractions to values whose side effects at other wavelengths project onto whichever template is being fitted. This is the quantitative counterpart of a finding made repeatedly in the literature \citep{Wakeford2019,Garcia2022,Lim2023,Iyer2023,RackhamDeWit2024,Espinoza2025}, that current model spectra of ultracool photospheres are not accurate enough for this purpose. The full stellar-model results, including its own null test, are given in Appendix~\ref{app:tls}.


\section{Application to TRAPPIST-1\,c}
\label{sec:application}

With the null distribution measured on \Tb\ (Section~\ref{sec:null}) and the response on the four \Tc\ epochs checked (Section~\ref{sec:injrec}), the framework is calibrated, and we now apply it to \Tc\ itself.

\subsection{No evidence for an atmospheric signal}
\label{sec:app_null}

For all three templates the fit to the real \Tc\ data prefers no signal. The evidence is negative in every case ($\dlnz = -2.5$, $-0.1$, and $-0.9$ for \co, water, and methane), and the recovered amplitudes are consistent with what the airless \Tb\ produces (Table~\ref{tab:null}). \Tc\ lies 0.7$\sigma$ below the null mean for \co, 0.4$\sigma$ above for water, and 1.2$\sigma$ above for methane, and in each case a substantial fraction of the airless subsets returns a larger value (72, 33, and 13 per cent).

Methane, at 1.2$\sigma$, deserves a closer look. Two facts argue against a planetary origin. Repeating the fit on the 2--5\um\ range alone, which keeps the strongest methane bands but excludes the most contaminated wavelengths, shows no excess (Appendix~\ref{app:red}). If the offset were planetary methane, the range containing its strongest bands should show it at least as clearly. And the airless \Tb\ shows no methane floor, so the offset is something specific to the four \Tc\ epochs rather than a persistent feature of the star (Section~\ref{sec:injrec_ch4}). We therefore make no claim of a detection, while noting that a methane-shaped signal of a few tens of ppm cannot be excluded by these data. For water, the $+70$\,ppm returned on \Tc\ is consistent with a single draw from the stellar floor distribution, and any planetary contribution can only be bounded, which we do next.

\subsection{Upper limits on the amplitude of planetary features}
\label{sec:app_limits}

Turning the posteriors into upper limits on a planetary feature takes two steps. The first is a physical constraint. The fit is allowed to return negative amplitudes, because the calibration has to be able to measure an offset of either sign, but a planetary absorption feature cannot be negative. When we convert a posterior into a limit on the planetary amplitude $A_{\rm p}$, we therefore discard the negative part of the distribution and read off the 95th percentile of what remains. This matters when the posterior median is negative, as it is for \co. Without the truncation the limit would be tighter than the data justify. The second step involves a choice, namely how much of the \Tb\ null distribution to use. Applying it to \Tc\ assumes that the four \Tc\ epochs are draws from the same distribution as random sets of four \Tb\ epochs (Section~\ref{sec:disc_limitations}), and we do not want the headline number to rest entirely on that assumption. We therefore give three limits of increasing assumption (Table~\ref{tab:limits}), all computed numerically from the posterior samples of the amplitude from the \Tc\ fit and the 100 null values from \Tb, with no Gaussian shapes assumed. The statistical-only limit uses the \Tc\ posterior samples as they are. The floor-scatter limit subtracts from each sample a value drawn from the \Tb\ null distribution with its mean removed, so it allows for the star adding a signal of either sign without assuming what its mean is. The floor-marginalized limit subtracts a value drawn from the null distribution as measured, mean included. In each case the negative results are discarded and the 95th percentile of the rest is the limit. Because the 100 null values come from only 17 distinct epochs, the null mean is itself uncertain by about 15\,ppm (Appendix~\ref{app:validation}), and this is included as an additional random shift wherever the mean enters.

\begin{table}
\centering
\caption{95 per cent upper limits (ppm) on the peak-to-trough amplitude of planetary template-shaped signals in \Tc's transmission spectrum, with the physical prior $A_{\rm p} \ge 0$. The three columns use progressively more of the \Tb\ null distribution: none (statistical-only), its scatter but not its mean (floor-scatter), and both (floor-marginalized), as described in Section~\ref{sec:app_limits}. The headline value for each molecule is in bold. Without the positivity prior the statistical-only (floor-marginalized) values are 50, 127, and 129 (65, 98, and 167)\,ppm for the three rows in order. The corresponding values for the 2--5\um\ range are given in Appendix~\ref{app:red}.}
\label{tab:limits}
\footnotesize
\setlength{\tabcolsep}{4pt}
\begin{tabular}{lccc}
\toprule
Template & Stat.-only & Scatter & Marg. \\
\midrule
\co\ full    & 70  & 93  & \textbf{90}  \\
\water\ full & \textbf{127} & 155 & 108 \\
CH$_4$ full  & 133 & 165 & \textbf{173} \\
CH$_4$ red   & 97  & 141 & \textbf{138} \\
\bottomrule
\end{tabular}
\end{table}

Which limit is the headline depends on the molecule. For \co\ and methane the null means on \Tb\ are consistent with zero, so subtracting the floor shifts the posterior little and mainly widens it by the epoch-to-epoch scatter, and the floor-marginalized values are the headline, $A_{\rm p} < 90$\,ppm for \co\ and $< 173$\,ppm for methane. In words, a planetary \co\ feature larger than 90\,ppm from peak to trough is ruled out at 95 per cent confidence, and likewise a methane feature larger than 173\,ppm. The methane limit is the weakest of the three because it inherits the offset discussed above. For water the headline is the statistical-only limit, $A_{\rm p} < 127$\,ppm, for the following reason. The amplitude measured on \Tc, $+70$\,ppm, is the sum of a stellar part and a planetary part, and the one thing \Tb\ establishes beyond doubt is that the stellar part is positive. The airless planet returns a water-like signal of $+55 \pm 34$\,ppm, above zero in 95 per cent of the subsets. The planetary part therefore cannot be larger than what is measured, and the 95th percentile of the measured amplitude, 127\,ppm, is a valid upper limit on the planet that needs no assumption about how large the stellar part is. Subtracting the \Tb\ floor gives a tighter value, 108\,ppm, but that assumes the stellar part on the \Tc\ epochs is the same as on random \Tb\ epochs, so we quote it as a refinement rather than as the headline. These amplitude limits are the result of this paper.

\subsection{What the amplitude limits mean for atmospheres}
\label{sec:pressure_mapping}

Up to this point we have worked in amplitudes only. That was deliberate. An amplitude makes no assumption about which atmosphere, at what temperature, surface pressure, or composition, would produce a feature of a given size, and it is the quantity our calibration actually controls. Translating it into statements about atmospheres requires models, and models carry assumptions that the data do not test. In this section we make that translation, cautiously, and say where the caveats lie.

The link between the two is the amplitude the fit would return for a real atmosphere. The fit can only scale the fixed template up or down, so for a model spectrum of a different shape it returns the amplitude at which the template best matches that spectrum, with each wavelength bin counted in proportion to how precisely the real \Tc\ data measure it (a weighted least-squares fit, ignoring the GP). We call this the equivalent amplitude. It is not the same as the peak-to-trough height of the model spectrum itself. It can be smaller or larger, depending on how the shape of the model differs from that of the template in the bins that carry most of the weight. We compute it from \textsc{platon} model spectra of clear, isothermal atmospheres with a solid surface at the equilibrium temperature of 342\,K \citep{Gillon2017}, and check below how the numbers move at 300 and 500\,K.

\begin{figure*}
\centering
\figfile{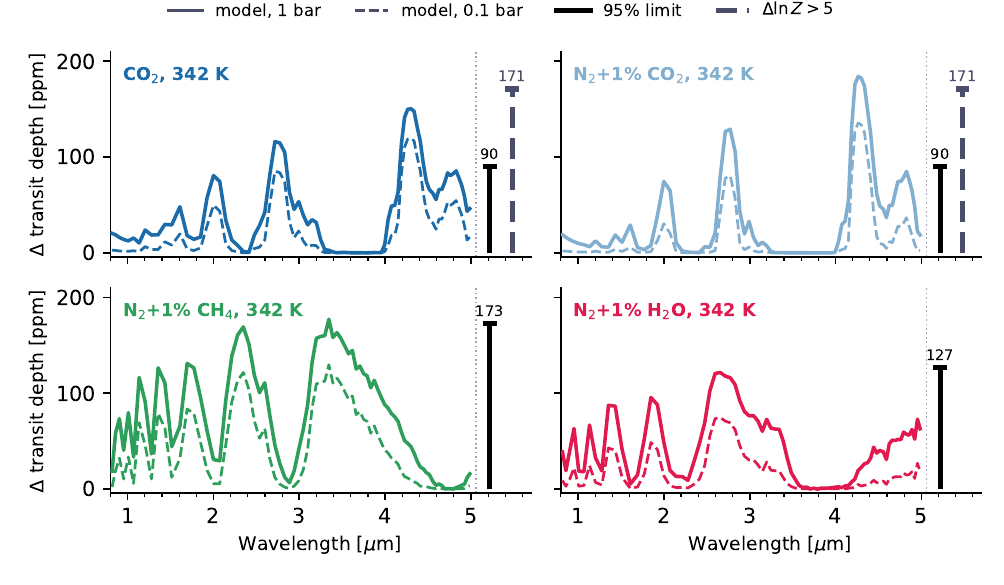}{0.9\textwidth}
\caption{Model transmission spectra of the four representative atmospheres of Table~\ref{tab:models} on \Tc\ at 342\,K, on the wavelength grid of the data and relative to the lowest point of each spectrum, at surface pressures of 1\,bar (solid) and 0.1\,bar (dashed). The vertical bars show the 95 per cent limit on the planetary amplitude for that molecule (\co\ 90\,ppm, CH$_4$ 173\,ppm, \water\ 127\,ppm) and, in the \co\ panels, the four-transit strong-detection threshold of 171\,ppm. The bars are to be compared with the equivalent amplitudes of Table~\ref{tab:models}, which are what the fit would return for these spectra, rather than with the height of the curves. The two differ because the fit matches a fixed template shape to the spectrum with the data weights. The peak-to-trough heights of the curves at 0.1 and 1\,bar are 118 and 149\,ppm for pure \co, 135 and 184\,ppm for N$_2$ with 1 per cent \co, 128 and 177\,ppm for N$_2$ with 1 per cent CH$_4$, and 80 and 128\,ppm for N$_2$ with 1 per cent \water, so the equivalent amplitude is below the curve height for the \co\ atmospheres and above it for the methane atmosphere.}
\label{fig:models}
\end{figure*}

Which surface pressures to compare at is not a free choice. For a planet with a solid surface, the amplitude is zero as long as the atmosphere is too thin to be opaque at any wavelength. It becomes non-zero once the centres of the strongest bands are opaque, and from there it grows only with the logarithm of the surface pressure, because each additional scale height of atmosphere raises the opaque level by the same height. It stops growing once the wavelengths between the bands are opaque as well, and beyond that point transmission spectroscopy cannot tell how much more atmosphere lies below. For the atmospheres in Table~\ref{tab:models}, a factor of ten in surface pressure, from 0.1 to 1\,bar, changes the equivalent amplitude by only 42 to 67\,ppm, and a change of 20\,ppm in a limit moves the pressure at which a model crosses it by a factor of two to three. An amplitude therefore constrains the presence of a thick atmosphere well and its surface pressure hardly at all, and inverting a limit into an excluded pressure would mean reading a nearly flat curve backwards. We do not do that. Instead we compare our limits with the amplitudes that four representative atmospheres would produce at two surface pressures, 0.1 and 1\,bar (Fig.~\ref{fig:models}; Table~\ref{tab:models}): pure \co, and N$_2$-dominated atmospheres with 1 per cent of \co, of CH$_4$, or of \water. These are pressures at which the models are on solid ground. What happens below 0.1\,bar, where the model spectra consist of narrow line cores and the projection onto a 1-bar template is stretched furthest, is left to dedicated modelling work.

\begin{table}
\centering
\caption{Equivalent amplitudes of the four representative clear, isothermal atmospheres of Fig.~\ref{fig:models} on \Tc\ at surface pressures of 0.1 and 1\,bar and the equilibrium temperature of 342\,K. The equivalent amplitude is what the amplitude fit would return for that atmosphere (Section~\ref{sec:pressure_mapping}). The equivalent amplitude is to be compared with the 95 per cent limits of Table~\ref{tab:limits} (\co\ 90\,ppm, CH$_4$ 173\,ppm, \water\ 127\,ppm) and, for \co, with the four-transit strong-detection threshold of 171\,ppm. The last two columns give the outcome of that comparison, for 0.1 and 1\,bar respectively. The detectability comparison is made only for \co, because the $\dlnz > 5$ thresholds for water and methane on the \Tc\ epochs are offset-assisted (Sections~\ref{sec:injrec_h2o} and \ref{sec:injrec_ch4}) and do not measure the sensitivity to a planetary signal. The \co\ rows are the robust ones, since the methane limit is offset-assisted and the water limit rests on the sign of the stellar floor (Section~\ref{sec:app_limits}).}
\label{tab:models}
\footnotesize
\setlength{\tabcolsep}{4pt}
\begin{tabular}{lcccc}
\toprule
 & \multicolumn{2}{c}{$A_{\rm equiv}$ [ppm]} & Above limit & Detectable \\
\cmidrule(lr){2-3}
Composition & 0.1\,bar & 1\,bar & 0.1 / 1\,bar & 0.1 / 1\,bar \\
\midrule
Pure \co\           & 103 & 145 & yes / yes & no / no \\
N$_2$ + 1\% \co\    &  87 & 154 & no / yes  & no / no \\
N$_2$ + 1\% CH$_4$  & 179 & 232 & yes / yes & -- \\
N$_2$ + 1\% \water\ &  74 & 137 & no / yes  & -- \\
\bottomrule
\end{tabular}
\end{table}

The comparison supports the following statements, which we regard as weak constraints rather than measurements. A pure \co\ atmosphere would produce 145\,ppm at 1\,bar and 103\,ppm at 0.1\,bar, both above our 95 per cent limit of 90\,ppm, so clear pure-\co\ atmospheres of 0.1\,bar or more are disfavoured. An N$_2$ atmosphere with 1 per cent \co\ would produce 154\,ppm at 1\,bar, above the limit, and 87\,ppm at 0.1\,bar, just below it. Neither atmosphere reaches the strong-detection threshold of 171\,ppm at either pressure, so four transits could disfavour them but could not have detected them, which is the difference between exclusion, set by the 95 per cent limit, and detectability, set by the $\dlnz > 5$ threshold. For methane, the limit of 173\,ppm inherits the epoch-specific offset and is therefore conservative. An N$_2$ atmosphere with 1 per cent CH$_4$ would produce 232 and 179\,ppm at 1 and 0.1\,bar, both above it. For water, N$_2$ with 1 per cent water would produce 137\,ppm at 1\,bar, above the 127\,ppm limit, and 74\,ppm at 0.1\,bar, below it. This is the least robust of the statements, because the water limit rests on the sign of the stellar floor. All of these numbers assume the equilibrium temperature. At 300\,K every amplitude in Table~\ref{tab:models} is 13 to 21 per cent smaller and three of the comparisons change sign: pure \co\ at 0.1\,bar drops to 82\,ppm and N$_2$ with 1 per cent CH$_4$ at 0.1\,bar to 143\,ppm, both below their limits, so those two statements then hold only from 1\,bar, and N$_2$ with 1 per cent water at 1\,bar drops to 119\,ppm, below the water limit. At 500\,K every amplitude is 1.5 to 2 times larger, and the two \co\ atmospheres at 1\,bar would then have exceeded the strong-detection threshold.

Thermal emission constrains \Tc\ from the other end. The 15\um\ eclipse of \citet{Zieba2023} and the 15\um\ phase curves of \citet{Gillon2026} disfavour thick \co\ atmospheres through the heat they would carry to the nightside and the \co\ absorption at 15\um, and the phase curves leave either an airless surface or a tenuous, greenhouse-poor O$_2$-dominated atmosphere. Transmission rules out thick \co\ atmospheres through the size of the 4.3\um\ feature they would produce. Together the two leave four kinds of scenario: no atmosphere, a thin one, an aerosol-flattened one, or a non-absorbing one. O$_2$- or N$_2$-dominated atmospheres with \co\ behave like the N$_2$+\co\ row of Table~\ref{tab:models}, with a smaller amplitude for a smaller \co\ content, whereas a pure N$_2$ or O$_2$ atmosphere without \co\ or CH$_4$ is not constrained at all by our data.

The main caveats are these. The models are clear and isothermal at the equilibrium temperature, so a cooler atmosphere gives smaller amplitudes and weaker statements, as the 300\,K values above show, and aerosols would flatten the spectrum and weaken every statement. We also note that the projection onto the 1-bar template ignores how the GP responds to a spectrum of a different shape, such as that of a thinner atmosphere, so the amplitude the fit would recover for such an atmosphere could be somewhat smaller than the projected one. The limits rely on the \Tb\ measurements less than one might expect. The water limit uses only the fact that the stellar contribution is positive, not its size, and the \co\ and methane limits use only the epoch-to-epoch scatter measured on \Tb, because the null means for those two molecules are consistent with zero. Finally, the limits apply to features shaped like our three templates. An atmosphere whose spectrum resembles none of them is not constrained.

\subsection{The cost of contamination}
\label{sec:app_whitenoise}

The white-noise test of Appendix~\ref{app:validation} lets us say what the contamination cost. In that test the four \Tc\ spectra are replaced by synthetic ones that keep the real wavelength grid and the real per-bin uncertainties but contain only Gaussian noise, so they represent the same observations without any stellar contamination. On those spectra the strong-detection threshold would have been 96\,ppm instead of 171\,ppm and the 95 per cent limit about 45 to 50\,ppm instead of 90\,ppm. In the terms of Table~\ref{tab:models}, that is the difference between ruling out thick \co\ atmospheres and detecting them. With the contamination-free threshold, a pure \co\ atmosphere of 0.1\,bar or more and an N$_2$ atmosphere with 1 per cent \co\ of 1\,bar or more would both have been strongly detected at the equilibrium temperature. The scaling study of the next section gives the exchange rate. The threshold on the \Tb\ archive reaches the contamination-free four-transit value at seven to eight transits (94 and 92\,ppm), so roughly twice the transits buy back the contamination-free sensitivity for the carbon species. For water there is no such exchange. The floor is independent of the number of transits, and from about eight transits on it is the star's water that becomes formally significant, not the planet's. For water the remedy is therefore not more transits. It is a measurement of the stellar floor, for which an airless sibling of the target is one route being pursued. Where no such measurement exists, a water detection on a rocky planet of a cool M dwarf should be treated as floor-limited rather than as a detection of the planet.


\section{How many transits are needed? An empirical scaling study}
\label{sec:scaling}

\begin{figure}
\centering
\figfile{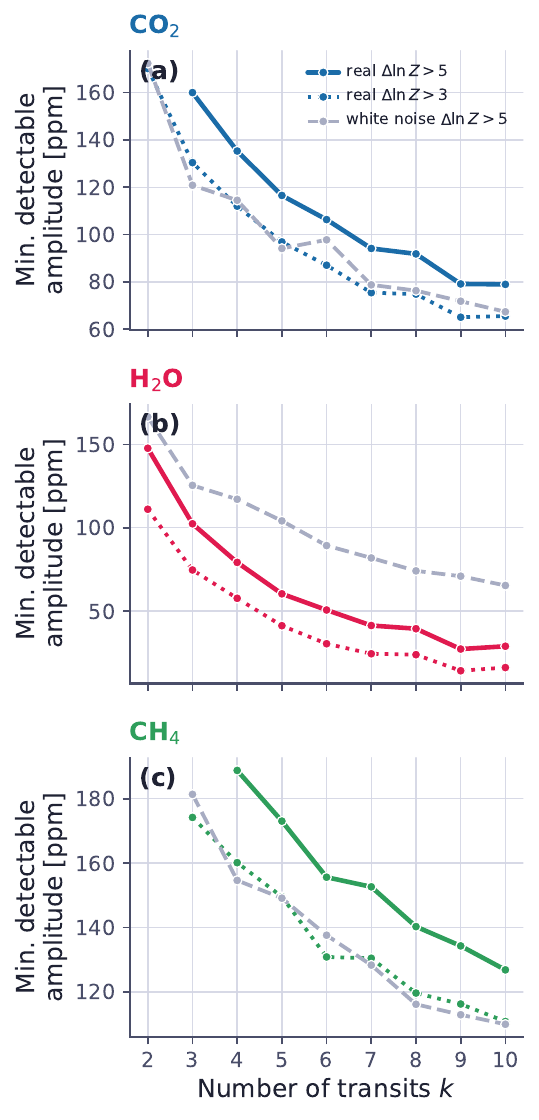}{\columnwidth}
\caption{Minimum detectable amplitude versus number of stacked transits $k$ on the \Tb\ archive, one panel per template, showing real-data thresholds at $\dlnz > 5$ (solid) and $\dlnz > 3$ (dotted), and the $\dlnz > 5$ threshold on white-noise realizations of the same subsets (grey dashed). Curves start at the smallest $k$ for which the threshold falls within the 200\,ppm injection grid. For \water\ the real threshold lies \emph{below} the white-noise one not because water is easier to detect, but because the stellar water bias inflates the recovered amplitude, so the evidence is higher while the result is biased (Section~\ref{sec:scaling}).}
\label{fig:scaling}
\end{figure}

When planning observations of such targets, the simplest expectation is that the smallest detectable feature shrinks as $1/\sqrt{k}$ when $k$ transits are combined, as it would if the noise were white and independent from epoch to epoch. Stellar contamination is neither, so the real return on additional transits could be better or worse than this, and the \Tb\ archive lets us measure it. For each $k$ from 2 to 10 we drew 50 random sets of $k$ transits from the 17 available, injected the three templates at 0 to 200\,ppm into each set, and recorded where the signals become detectable. To keep the number of parameters manageable we used the GPTLS variant with one length scale shared among visits and precomputed templates (Appendix~\ref{app:scaling_details}). Table~\ref{tab:scaling} and Fig.~\ref{fig:scaling} give the thresholds on the evidence criterion, Table~\ref{tab:scaling_post} those on the posterior criterion together with the detection probabilities of Section~\ref{sec:protocol}, and the same runs on white-noise realizations of the same subsets (Appendix~\ref{app:validation}) provide the contamination-free comparison.

\begin{table}
\centering
\caption{Empirical detection thresholds (ppm, peak-to-trough, full wavelength range) versus number of stacked transits $k$, from 50 random $k$-transit subsets of the 17-transit \Tb\ archive per $k$, with GPTLS. Thresholds are the injected amplitudes at which the mean $\dlnz$ across subsets crosses 5 and 3. ``$>$200'' indicates no crossing within the injection grid. One fit (\co, $k=2$, 150\,ppm) failed to converge and is excluded, so that grid point averages 49 subsets. The \water\ thresholds are lowered by the persistent stellar water floor (Section~\ref{sec:injrec_h2o}) and cannot be read as planetary-detection thresholds.}
\label{tab:scaling}
\scriptsize
\setlength{\tabcolsep}{2.5pt}
\begin{tabular}{ccccccc}
\toprule
 & \multicolumn{2}{c}{\co} & \multicolumn{2}{c}{\water} & \multicolumn{2}{c}{CH$_4$} \\
\cmidrule(lr){2-3}\cmidrule(lr){4-5}\cmidrule(lr){6-7}
$k$ & $\dlnz>5$ & $\dlnz>3$ & $\dlnz>5$ & $\dlnz>3$ & $\dlnz>5$ & $\dlnz>3$ \\
\midrule
2  & $>200$ & 170 & 148 & 111 & $>200$ & $>200$ \\
3  & 160 & 130 & 102 & 75 & $>200$ & 174 \\
4  & 135 & 112 & 79 & 58 & 189 & 160 \\
5  & 117 & 97 & 60 & 41 & 173 & 150 \\
6  & 106 & 87 & 51 & 31 & 156 & 131 \\
7  & 94 & 76 & 41 & 25 & 153 & 131 \\
8  & 92 & 75 & 40 & 24 & 140 & 120 \\
9  & 79 & 65 & 27 & 14 & 134 & 116 \\
10 & 79 & 66 & 29 & 16 & 127 & 111 \\
\bottomrule
\end{tabular}
\end{table}

With the scaling study we find three results.

First, the thresholds fall with $k$ almost exactly as they would for white noise. For \co\ the strong-detection threshold drops from 135\,ppm at $k = 4$ to 79\,ppm at $k = 10$, where $1/\sqrt{k}$ scaling would predict about 85\,ppm. The fitted exponent is $-0.60$ against $-0.56$ on white noise and $-0.50$ for pure averaging, a difference within the 10 to 20\,ppm uncertainty of the thresholds. The contamination does not change how the thresholds scale. It raises the \co\ threshold by 10 to 30 per cent at every $k$ (Table~\ref{tab:wn_scaling}), and the scatter of the recovered amplitude from one subset to another is the same on real and white-noise data. In other words, each added epoch is an independent draw of the contamination, and the GP averages it down like noise. Methane's thresholds are higher than those of \co\ on this archive, on white noise as well as on the real data, so the difference is a property of the methane template on these error bars rather than of the contamination. Water's steep fall is not a gain in sensitivity but rather the stellar floor becoming detectable.

Second, the evidence criterion has a floor. At this per-transit precision a 50\,ppm \co\ signal never reaches $\dlnz = 3$ for $k \le 10$. The Occam penalty of Section~\ref{sec:inference} keeps the evidence near zero even when the posterior recovers the amplitude accurately, and the white-noise runs show the same behaviour, so it is a property of the statistic, not of the contamination. The floor is set by the ratio of the amplitude to the posterior width, so it recedes in proportion to the per-bin precision of the data, and it is not a limit on the amplitude the method can reach. The posterior criterion has no such floor, and its 3$\sigma$ threshold reaches 61\,ppm at $k = 10$ for \co\ (Table~\ref{tab:scaling_post}). For programme design we therefore also report the amplitude at which 90 per cent of subsets yield $\dlnz > 5$, which captures the luck of which epochs one happens to observe. A 200\,ppm \co\ feature is strongly detected in 44, 90, and 98 per cent of two-, three-, and four-transit subsets.

\begin{table*}
\centering
\caption{Posterior-criterion thresholds and detection probabilities versus number of stacked transits $k$ (GPTLS, full wavelength range, 50 random $k$-transit subsets of the 17 \Tb\ transits). $z{>}3$: injected amplitude (ppm) at which the mean posterior significance $A_{\rm med}/\sigma_A$ crosses 3. $A_{90}$: injected amplitude (ppm) at which 90 per cent of subsets yield $\dlnz > 5$. FP: fraction of zero-injection subsets with $z > 2$ (for \water\ this is the stellar floor being detected). ``$>$200'': not reached within the injection grid.}
\label{tab:scaling_post}
\footnotesize
\setlength{\tabcolsep}{4pt}
\begin{tabular}{cccccccccc}
\toprule
 & \multicolumn{3}{c}{\co} & \multicolumn{3}{c}{\water} & \multicolumn{3}{c}{CH$_4$} \\
\cmidrule(lr){2-4}\cmidrule(lr){5-7}\cmidrule(lr){8-10}
$k$ & $z{>}3$ & $A_{90}$ & FP & $z{>}3$ & $A_{90}$ & FP & $z{>}3$ & $A_{90}$ & FP \\
\midrule
2  & 169 & $>$200 & 0.00 & 116 & $>$200 & 0.16 & $>$200 & $>$200 & 0.02 \\
3  & 125 & 200 & 0.02 & 73 & 160 & 0.22 & 175 & $>$200 & 0.00 \\
4  & 105 & 187 & 0.00 & 54 & 139 & 0.32 & 157 & $>$200 & 0.02 \\
5  & 90  & 175 & 0.02 & 38 & 125 & 0.46 & 146 & $>$200 & 0.00 \\
6  & 80  & 143 & 0.02 & 28 & 94  & 0.48 & 127 & 200 & 0.00 \\
7  & 71  & 138 & 0.00 & 21 & 87  & 0.66 & 125 & 196 & 0.00 \\
8  & 71  & 138 & 0.00 & 21 & 88  & 0.62 & 116 & 188 & 0.00 \\
9  & 61  & 122 & 0.00 & 11 & 75  & 0.82 & 112 & 182 & 0.00 \\
10 & 61  & 114 & 0.02 & 15 & 81  & 0.64 & 105 & 158 & 0.00 \\
\bottomrule
\end{tabular}
\end{table*}

Third, the templates hit different walls. \co\ and methane are limited by the statistics of the contamination throughout, with no false positives at any $k$. Water is limited by the floor. The spurious amplitude on airless subsets stays at 44 to 56\,ppm whatever $k$ is, while the contamination-free strong-detection threshold falls from 117\,ppm at $k = 4$ to 65\,ppm at $k = 10$, and the contamination-free moderate-evidence threshold reaches the floor at $k = 10$ (56\,ppm). Well before that the floor inflates the false-positive rates on the real data, from 2 per cent at $k = 4$ to 12 per cent at $k = 6$ and 10 to 14 per cent at $k = 9$ to 10 on the evidence criterion, and from 16 to more than 60 per cent on the posterior criterion (Table~\ref{tab:scaling_post}). Water searches around stars like TRAPPIST-1 are thus prone to floor-driven false positives from about six transits, and by ten transits the floor is as large as the contamination-free moderate-evidence threshold. Progress requires measuring the floor, for which the use of an airless companion, statistically as here or with back-to-back transits, is being pursued \citep{TJCI2024,Rathcke2025,Allen2026}.

We note that this scaling study comes with two caveats. The 50 subsets at each $k$ are drawn from only 17 epochs and overlap, which does not bias the thresholds but makes their uncertainties larger than $1/\sqrt{50}$ would suggest, and the shared length scale mildly restricts the GP (Appendix~\ref{app:scaling_details}). And the thresholds are peak-to-trough template amplitudes, and what they mean for a given atmosphere requires the models of Section~\ref{sec:pressure_mapping}. For \Tc\ this means that a 200\,ppm feature is strongly detected with three or more transits, whereas a clear pure-\co\ atmosphere at the equilibrium temperature, whose equivalent amplitude is 145\,ppm at 1\,bar and reaches at most 165\,ppm even at 10\,bar in our models, needs four typical transits and was not detectable at all with the actual four \Tc\ epochs.

\FloatBarrier


\section{Discussion}
\label{sec:discussion}

\subsection{The combined observational picture for TRAPPIST-1\,c}
\label{sec:disc_combined}

Four JWST datasets now constrain \Tc's atmosphere from different directions. The 15\um\ eclipse of \citet{Zieba2023} found a dayside temperature between the values expected for no and for full heat redistribution, and disfavoured thick \co-dominated atmospheres. The 15\um\ phase curves of \citet{Gillon2026} extended this to the nightside. \Tc\ has a dayside brightness temperature of $369 \pm 23$\,K, cooler than \Tb, and a nightside flux indistinguishable from that of the airless \Tb, which strongly disfavours atmospheres of 1\,bar or more with an efficient greenhouse effect and leaves either an airless planet with a more reflective surface than \Tb\ or a tenuous, greenhouse-poor O$_2$-dominated atmosphere. The NIRISS transmission spectra of \citet{Radica2025} excluded H$_2$-dominated atmospheres and disfavoured thick atmospheres rich in water, NH$_3$, or CO, but, lacking the 3--5\um\ region, could not constrain \co- or CH$_4$-rich compositions. This work closes that gap from the transmission side. For clear, isothermal atmospheres at the equilibrium temperature, pure \co\ atmospheres of 0.1\,bar or more, N$_2$-dominated atmospheres with 1 per cent \co\ of 1\,bar or more, and N$_2$ atmospheres with 1 per cent CH$_4$ of 0.1\,bar or more would all have produced features larger than our limits (Table~\ref{tab:models}), and at 300\,K the pure-\co\ and methane statements move to 1\,bar. These are exclusions rather than detections. No clear \co-bearing atmosphere at the equilibrium temperature could have been strongly detected with four transits, and where exactly the boundary lies in pressure depends strongly on the assumed temperature and on the models, which is why we state it at 0.1 to 1\,bar rather than as a precise value.

What survives is a short list. There may be no atmosphere. There may be a thin \co-bearing one (a Mars-like atmosphere is allowed, and thinner atmospheres are beyond what our models can be trusted to constrain). There may be a non-absorbing one, since pure N$_2$ or O$_2$ without \co\ or CH$_4$ is not constrained at all. Or there may be an aerosol-covered one, whose flat spectrum evades template-shaped searches. The tenuous O$_2$-dominated atmosphere that the phase curves leave open belongs to the third category, and our data constrain it only through whatever \co\ or CH$_4$ it carries (Table~\ref{tab:models}). This is consistent with the expectation that \Tc, given its cumulative XUV exposure, retains at most a tenuous secondary atmosphere \citep{ZahnleCatling2017,KrissansenTotton2023,Turbet2020}, although evolutionary models also allow thick O$_2$- or \co-dominated atmospheres \citep{Lincowski2018}, and with the emerging pattern across the inner TRAPPIST-1 planets \citep{Greene2023,Ducrot2025,Gillon2026,Lim2023,Radica2025}.

\subsection{The stellar water floor: a generic hazard for water searches around late M dwarfs}
\label{sec:disc_water}

The spurious water amplitude of $+55 \pm 34$\,ppm on airless \Tb\ shows that the transit light source effect of an M8V star imprints a static, planet-like water signature on transmission spectra, which survives any correction based on epoch-to-epoch variability. GP marginalization, epoch differencing, and parametric fits with epoch-varying covering fractions all remove only the variable part. Two features make it dangerous. It is a distribution rather than a constant, whose width of 34\,ppm is set by which epochs one observes and is itself the irreducible systematic for any water measurement, so a target's apparent water amplitude can only be bounded (Section~\ref{sec:app_limits}). And it grows more, not less, deceptive with data volume. As transits accumulate the statistical uncertainty shrinks below the floor and the floor itself becomes formally detectable, and on airless \Tb, 64 to 82 per cent of nine- and ten-transit subsets show a $>$2$\sigma$ posterior water signal (Section~\ref{sec:scaling}). Its size, several tens of ppm, is comparable to an Earth-like water signal for a temperate TRAPPIST-1 planet, so claimed water detections on planets of ultracool dwarfs should be regarded with suspicion unless the floor has been measured directly. \citet{Moran2023} provide an instructive analogue for an earlier-type M dwarf, where a tentative water feature on GJ~486\,b could equally be attributed to unocculted starspots. An airless companion of the same star offers a way to measure the floor, either statistically, as done here with the \Tb\ archive, or directly with back-to-back or simultaneous transits \citep{TJCI2024,Rathcke2025,Allen2026}. How completely the contamination transfers from one planet to another is still being established, and rests on the assumption discussed in Section~\ref{sec:disc_limitations}. Within that assumption, our measurement of the water floor gives the size of the effect to expect in other TRAPPIST-1 transmission spectra. By contrast, \co\ and methane are clean. The star contributes no persistent \co- or methane-like signal at our sensitivity, so the carbon species remain the most trustworthy molecules for atmosphere searches in this and similar systems, which is fortunate, since \co\ is also among the most physically expected heavy volatiles \citep{Lincowski2018,Turbet2020,WordsworthKreidberg2022}.

\subsection{Relation to previous work, and limitations}
\label{sec:disc_limitations}
\label{sec:disc_gp_vs_tls}

Our approach is closest to that of \citet{Espinoza2025} for \Te, who fit all epochs jointly with a GP per epoch and an atmospheric signal shared between them. We add two things. The \Tb\ null test converts the central untestable assumption of a single-planet analysis, that no contamination component persists across epochs, into a measured quantity, and it is what exposes the offset-only model, which passes every test on the science target and returns a $+56$\,ppm spurious \co\ signal on the airless planet. And the injection-recovery calibration on the real data, together with the white-noise validation, measures throughput, coverage, thresholds, and the cost of the contamination directly rather than assuming them. In the other direction, \citet{Espinoza2025} and \citet{Glidden2025} run full atmospheric retrievals under the GP marginalization, whereas we fit one template amplitude. The template approach buys calibratability at the cost of compositional detail, and the two are complementary. The parametric stellar-model correction, which we ran as a comparison, is biased by up to 204\,ppm with a sign that depends on the template and fails its own null test, in line with the known inaccuracy of ultracool-photosphere models at JWST precision \citep{Wakeford2019,Garcia2022,Lim2023,Iyer2023,RackhamDeWit2024}.

The limitations are the following. The constraints are template-shaped, so an atmosphere whose spectrum resembles none of the three templates, or a flat aerosol-dominated one, is not constrained. The model amplitudes of Table~\ref{tab:models} assume clear, isothermal atmospheres and a projection that ignores the GP's response to shape mismatch, and they are only trusted at the pressures where they are quoted (Section~\ref{sec:pressure_mapping}). Using the \Tb\ null for \Tc\ assumes that the statistics of the contamination are the same along the two transit chords and across the two sets of epochs. The near-identical impact parameters \citep{Agol2021} and the demonstrated similarity of the contamination seen by b and c within a single epoch \citep{Rathcke2025} support this, but a chord- or season-dependent component would not be captured. The 100 subsets are drawn from 17 epochs and overlap, so the tails of the null distribution are less well sampled than 100 independent draws would be (Appendix~\ref{app:scaling_details}). The residual cost of the contamination after GP marginalization is a factor of 1.2 in threshold for typical epochs and 1.8 for the flare-affected \Tc\ set, paid in posterior width rather than bias (Appendix~\ref{app:validation}). And the fixed-template approach is a mildly optimistic stand-in for a full retrieval, whose additional parameters would raise the evidence thresholds by perhaps a few tens of per cent.

\subsection{Recommendations for future observations}
\label{sec:disc_recommendations}

For \co-focused searches around mid-to-late M dwarfs with PRISM, on a star like TRAPPIST-1, three transits suffice for a strong detection of a 200\,ppm feature, four transits reach about 135\,ppm on typical epochs and 171\,ppm on the actual \Tc\ epochs, and nine to ten transits reach about 80\,ppm, below the amplitude of every 1\,bar atmosphere in Table~\ref{tab:models}. The thresholds fall as $1/\sqrt{k}$ provided the contamination is marginalized jointly across epochs. At \Tc-like precision the contamination-free four-transit \co\ threshold is about 96\,ppm, which the contamination raises by 20 per cent for typical epochs and 80 per cent for the flare-affected \Tc\ set. A 40\,ppm feature would need about 20 transits without contamination and 25 to 45 with it, and a target with twice the per-bin precision needs a quarter as many. In short, GP marginalization with empirical calibration turns TRAPPIST-1's stellar contamination from an unquantified systematic into a known cost, roughly a factor of two in transits for the carbon species, while water remains floor-limited unless the floor is measured, for which an airless companion of the same star, observed back-to-back or, as here, used as a statistical control sample, is the most promising avenue to explore \citep{TJCI2024,Rathcke2025,Allen2026}.

Most systems do not offer an airless sibling. Without one, injection-recovery on the target still calibrates throughput, coverage, and the false-positive rate for an added signal, but it cannot reveal a persistent floor. We recommend four mitigations. First, rely on prior physics, since the photosphere of a late M dwarf plausibly has no persistent \co- or methane-like features but certainly has water, and treat any water claim as floor-limited. Second, use model-independent variability metrics of the kind in Section~\ref{sec:variability} as a proxy for the size of a possible floor. Third, cross-check any signal against the part of the bandpass where the contamination is weakest, as we do for methane in Section~\ref{sec:app_null}. Fourth, report the ratio of the target's zero-injection amplitude to its posterior width alongside the evidence, so that readers can judge a claim on both criteria. More broadly, any claimed molecular detection on a planet of an active M dwarf should come with an injection-recovery calibration on the actual data and, where the architecture permits, a null test on an airless sibling. Those were the two tests that exposed a $+56$\,ppm spurious \co\ signal that every target-only test had missed.


\section{Conclusions}
\label{sec:conclusions}

We have presented GPTLS, an empirically calibrated Gaussian-process framework for atmosphere detection in the presence of stellar contamination, and applied it to four JWST/NIRSpec PRISM transits of \Tc\ from GO~2420, the first multi-epoch transmission spectroscopy of this planet covering the 3--5\um\ region. Our main conclusions are:

\begin{enumerate}
\item \textbf{An airless neighbour calibrates the search.} A null distribution built from 100 four-transit subsets of 17 archival PRISM transits of the airless \Tb\ provides the first empirical measurement of the spurious signals that stellar contamination produces in the TRAPPIST-1 system. It turns the assumption underlying single-planet analyses, that no contamination persists across epochs, into a measured quantity, and it exposes a $+56$\,ppm spurious \co\ signal for a model without a GP that passes every injection test on the science target, so target-only calibration cannot reveal a persistent floor.

\item \textbf{The star has a water floor, and \co\ and methane are clean.} For \co\ and methane the framework is unbiased, with null distributions centred on zero. For water it is not. The airless \Tb\ returns $+55 \pm 34$\,ppm, a stellar water floor from the transit light source effect, which no epoch-differential method can remove and which becomes formally detectable as transits accumulate (false-positive rates of 10 to 14 per cent at $\dlnz > 3$ and above 60 per cent at 2$\sigma$ posterior significance for nine to ten transits).

\item \textbf{\co\ is recovered faithfully, at a known cost.} On the four \Tc\ epochs, injected \co\ signals are recovered one-for-one with honest uncertainties, and a feature needs 146 and 171\,ppm peak-to-trough to reach $\dlnz > 3$ and 5. The same epochs without contamination would give 96\,ppm, so the contamination costs them a factor of 1.8, against 1.2 for typical epochs of the archive. Methane recovery carries a $+46$\,ppm offset specific to the \Tc\ epochs, from its bands shortward of 2\um\ where the contamination is strongest.

\item \textbf{No atmospheric signal on \Tc.} The \Tc\ data show no evidence for an atmospheric signal. The evidence is negative for all three templates and every recovered amplitude is consistent with the airless null. The largest deviation, $+1.2\sigma$ ($p = 0.13$) for methane, disappears when the fit is restricted to 2--5\um\ and has no counterpart on \Tb, and the water amplitude lies $0.4\sigma$ above the airless mean.

\item \textbf{Thick \co\ atmospheres are ruled out, but could not have been detected.} With the physical prior $A_{\rm p} \ge 0$, the 95 per cent limits on planetary features are 90\,ppm for \co, 173\,ppm for methane, and 127\,ppm for water. For clear, isothermal atmospheres at the equilibrium temperature, pure \co\ atmospheres of 0.1\,bar or more, N$_2$-dominated atmospheres with 1 per cent \co\ of 1\,bar or more, and N$_2$ atmospheres with 1 per cent CH$_4$ of 0.1\,bar or more would have exceeded these limits (at 300\,K the pure-\co\ and methane statements move to 1\,bar). Where the boundary lies below that depends on the models and is not constrained by this work. The limits complement the NIRISS and emission constraints \citep{Radica2025,Zieba2023,Gillon2026}. No clear \co-bearing atmosphere at the equilibrium temperature could have been strongly detected with four transits, and roughly twice the transits would buy back the contamination-free sensitivity.

\item \textbf{The parametric correction fails.} A parametric stellar-model correction with two SPHINX components is biased by up to 204\,ppm with a sign that depends on the template, never contains the injected value in its credible intervals, and fails its own null test. Calibrated GP marginalization is the method of choice for such stars at present.

\item \textbf{Thresholds fall as $1/\sqrt{k}$, and water hits the floor.} On the \Tb\ archive, the \co\ detection threshold falls with the number of transits close to, and slightly faster than, $1/\sqrt{k}$, reaching about 80\,ppm at $k = 9$ to 10 with no false positives. The contamination raises the threshold by 10 to 30 per cent at every $k$ but does not change how it scales. A 200\,ppm \co\ feature is strongly detected in 44, 90, and 98 per cent of two-, three-, and four-transit programmes. Water searches become floor-limited beyond about six transits, and features below about 50\,ppm never reach $\dlnz = 3$ for $k \le 10$ because of the Occam penalty of the evidence, although the posterior continues to resolve them.

\item \textbf{The GP costs nothing without contamination.} White-noise simulations on the same error bars confirm that the framework is unbiased, with unit throughput, calibrated posteriors, and no false positives, and that the GP costs nothing when there is no contamination to model.
\end{enumerate}

The combination demonstrated here, a null test on an airless companion, injection-recovery calibration on the real data, and flexible marginalization of the contamination, is a transferable recipe for credible atmosphere searches on rocky planets of active M dwarfs, and the empirical scaling relations give concrete guidance for the deeper programmes that planets like \Tc\ will require.

\section*{Acknowledgements}
This work is based on observations made with the NASA/ESA/CSA James Webb Space Telescope. The data were obtained from the Mikulski Archive for Space Telescopes at the Space Telescope Science Institute, which is operated by the Association of Universities for Research in Astronomy, Inc., under NASA contract NAS 5-03127 for JWST. These observations are associated with programmes GO~2420, GO~1981, GO~6456, GO~9256, and DD~12492.

\section*{Data Availability}
The JWST data used in this work are publicly available from the Mikulski Archive for Space Telescopes (MAST). The \Tc\ observations were obtained under GO programme 2420, and the \Tb\ observations under programmes GO~1981, GO~2420, GO~6456, GO~9256, and DD~12492.

\bibliographystyle{mnras}
\bibliography{references}

\appendix

\section{Per-visit light-curve fits and posteriors}
\label{app:pervisit}
The white light-curve posteriors of the four \Tc\ epochs are given in Table~\ref{tab:priors_lightcurve}, and corner plots of the joint GP fits and the native-resolution spectra are provided with the analysis products (Data Availability).

\section{Full results of the stellar-model approach}
\label{app:tls}

The stellar-model approach is specified in Section~\ref{sec:tls_model}. It has a cool spot component at $T_{\rm spot} = 2360$\,K and a flare component approximated as a 5000\,K black body, with per-visit covering fractions $f_{{\rm spot},v} \in [0, 0.6]$ and $f_{{\rm flare},v} \in [0, 0.2]$, photosphere and spot spectra from the SPHINX grid \citep{Iyer2023} with $T_{\rm phot} = 2560$\,K, and otherwise identical data, templates, priors, and sampler settings to GPTLS. All stellar-model fits (the \Tc\ injection grids and the 100-subset control sample, for the three templates, over the full range and over the 2--5\um\ range of Appendix~\ref{app:red}) completed.

The injection-recovery statistics are given in the stellar-model rows of Table~\ref{tab:injrec} and the null-test results in those of Table~\ref{tab:null}. Three features summarize the model's behaviour. First, the bias is large and template-dependent in sign, $-204$\,ppm for \co\ and $-178$\,ppm for \water, but $+57$\,ppm for CH$_4$. Because the two fixed-temperature components cannot reproduce the observed contamination morphology, the fit drives the covering fractions to values whose chromatic residuals project onto each template in whichever direction that template's bands correlate with them. Second, throughput is slightly below unity ($\alpha = 0.96$--0.97) and coverage is 0 per cent for all three templates, and spurious $\dlnz > 3$ detections occur at zero injection for \co\ and \water. Third, and decisively, the model does not produce a consistent null between the two planets. On the airless \Tb\ subsets the stellar-model null distributions are themselves displaced, by $-53$\,ppm for \co, $-69$\,ppm for \water, and $+25$\,ppm for CH$_4$, and the \Tc\ amplitudes it recovers are $4.1\sigma$ (\co) and $2.0\sigma$ (\water) outliers from those distributions. Over the 2--5\um\ range its methane null is displaced to $+99$\,ppm, so that 59 per cent of the airless subsets exceed $z = 2$ and 28 per cent exceed $\dlnz = 3$ (Table~\ref{tab:null}), so a majority of airless four-transit sets would have been called a methane detection. A parametric correction that yields different biases on two airless-or-nearly-airless planets of the same star cannot be calibrated by either.

\section{Simplifications and validation of the transit-scaling study}
\label{app:scaling_details}

\paragraph*{Shared length scale.} The scaling runs use the reduced GP variant of Section~\ref{sec:gp_model} in which a single Mat\'ern-3/2 length scale is shared among visits, so that the parameter count grows as $3k + 2$ (14 at $k = 4$, 32 at $k = 10$) rather than $4k + 1$ (17 and 41). Templates were precomputed with \textsc{platon} and read from file, with identical physics to the main runs. We validated the reduced model against the full per-visit-$\ell$ model at $k = 4$, where both were run on the same archive. The subset-to-subset scatter of the recovered zero-injection \co\ amplitude is 30\,ppm in the scaling run against $\sigma_{\rm b} = 31$\,ppm in the 100-subset control run with per-visit length scales, and the mean amplitudes agree to within a few ppm. Evidence values from independent refits of the same configuration reproduce to $\sim$0.5--1 in $\dlnz$, so thresholds are quoted rounded to $\approx$10\,ppm and $\dlnz$ differences below $\sim$1 are not interpreted.

\paragraph*{Overlap of subsets.} Two random $k$-subsets of 17 epochs share on average a fraction $k/17$ of their epochs, which suppresses the observed subset-to-subset scatter by $\approx\sqrt{1 - k/17}$ but does not bias the mean threshold. It does reduce the number of effectively independent subsets, so threshold uncertainties exceed the naive $1/\sqrt{50}$ expectation. The raw scatter of the recovered \co\ amplitude at zero injection is 49, 31, 30, 25, 20, 19, 19, 14, and 14\,ppm for $k = 2, \ldots, 10$ (a raw exponent of $-0.74$). After overlap correction it is 52, 34, 34, 29, 25, 25, 26, 20, and 22\,ppm, with exponent $-0.51$, i.e.\ $1/\sqrt{k}$. At $k = 4$ the \Tc\ posterior half-width ($\sigma_A \approx 41$\,ppm) exceeds the corrected scatter (34--36\,ppm) by a factor $\approx$1.1--1.2.

\paragraph*{Completeness and sampling behaviour.} Of the 6\,750 GP scaling fits, one failed to converge (\co, $k = 2$, 150\,ppm, the subset of epochs 2024-12-10 and 2025-07-11). The dynamic sampler oscillated between the signal and no-signal modes for more than 24\,h. It is excluded, and that grid point averages 49 subsets, and all other fits completed. We found that sampling difficulty varies between epochs, and subsets containing certain \Tb\ visits required substantially longer run times, indicating more complex posterior geometries for the more active epochs.

\section{The flare in the 2023 November 8 epoch}
\label{app:flare}

Fig.~\ref{fig:flare} shows the flare of the 2023 November 8 visit and its effect on the transmission spectrum. In the white light curve (panel a) the flare falls during egress, and in the native-resolution H$\alpha$ channel (panel b) it stands out as a sharp emission spike. Panel (c) compares the transmission spectrum used in this work, in which nothing is masked, with one from an earlier version of the reduction in which the flare interval was excluded from the light-curve fit. The two differ mainly by a near-uniform shift in transit depth across the band, which is expected when part of the egress is removed from the fit and which the per-visit depth offset of our joint model absorbs by construction (Section~\ref{sec:gp_model}). Beyond that shift the differences are small compared with the epoch-to-epoch structure of Section~\ref{sec:variability} and largest in the bluest bins, where the flare contributes most and where the per-visit GP is left to absorb the residual. Because the masked spectrum comes from a previous version of the pipeline, part of the difference may also reflect reduction changes unrelated to the flare.

\begin{fixedfigure}{The 2023 November 8 flare. (a) Normalised white light curve with the masked interval shaded, where the red points are those excluded in the masked reduction. (b) Normalised flux in the native-resolution H$\alpha$ channel, in which the flare appears as an emission spike during egress. (c) The transmission spectrum used in this work (grey, no masking) compared with the spectrum from an earlier reduction in which the flare interval was masked (red).}{fig:flare}
\figfile{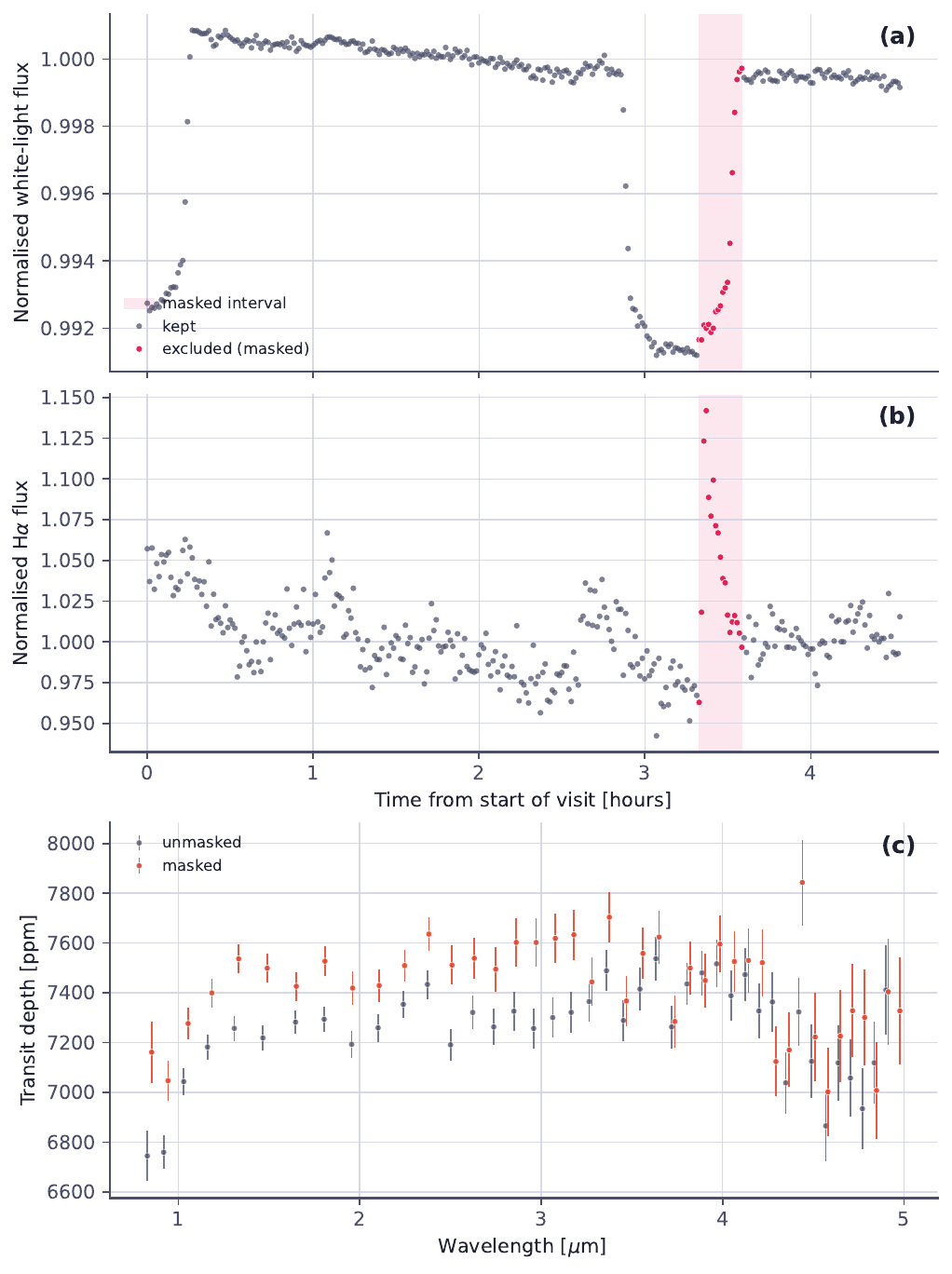}{\columnwidth}
\end{fixedfigure}

\needspace{18\baselineskip}
\section{TRAPPIST-1\,b observation log}
\label{app:t1b_log}
\begin{fixedtable}{The 17 archival NIRSpec PRISM transits of \Tb\ used for the control sample and the transit-scaling study. The 2023-12-11 visit is a TRAPPIST-1\,h observation from GO~1981 during which \Tb\ also transited. The 2024-07-09 visit is the double transit of \Tb\ and \Tc\ from our own programme \citep{Rathcke2025} and is the one epoch shared with the science dataset (Section~\ref{sec:obs_t1b}). The remaining visits are from the TRAPPIST-1\,b and e programmes GO~6456 and GO~9256 and from the DD programme 12492.}{tab:t1b_log}
\footnotesize
\setlength{\tabcolsep}{4pt}
\begin{tabular}{lll@{\hspace{12pt}}lll}
\toprule
\# & Date (UT) & Programme & \# & Date (UT) & Programme \\
\midrule
1 & 2023-12-11 & GO 1981  & 10 & 2025-07-11 & GO 9256 \\
2 & 2024-07-05 & GO 6456  & 11 & 2025-07-18 & GO 9256 \\
3 & 2024-07-09 & GO 2420  & 12 & 2025-10-29 & GO 9256 \\
4 & 2024-07-11 & GO 6456  & 13 & 2025-11-05 & GO 9256 \\
5 & 2024-12-10 & GO 6456  & 14 & 2025-11-17 & GO 9256 \\
6 & 2025-06-05 & GO 6456  & 15 & 2025-11-23 & GO 9256 \\
7 & 2025-06-11 & GO 6456  & 16 & 2025-12-11 & DD 12492 \\
8 & 2025-06-17 & GO 6456  & 17 & 2026-06-06 & GO 9256 \\
9 & 2025-06-29 & GO 9256  &    &            &         \\
\bottomrule
\end{tabular}
\end{fixedtable}

\section{White-noise validation}
\label{app:validation}

To separate the pipeline-and-inference behaviour of the framework from the astrophysical contamination, we repeated the analysis on synthetic spectra built from every real spectrum (17 \Tb\ and 4 \Tc). The wavelength grid and per-bin uncertainties were retained, and the depths replaced by the visit's weighted-mean depth plus Gaussian noise drawn from the uncertainties (one realization per visit, fixed seed), with no correlated structure of any kind. File names were retained so that the random subset draws are the same visit combinations as in the real scaling study, and the same templates, priors, sampler settings, shared-$\ell$ variant, and injection grids were used. Two sets of runs were made, the four-\Tc-visit analogue of Section~\ref{sec:injrec} (0--500\,ppm, three templates, full range, per-visit length scales) and the scaling analogue of Section~\ref{sec:scaling} ($k = 2$--10, 25 subsets per $k$, 0--200\,ppm, three templates, GPTLS and offset-only).

The four-visit results are given in the white-noise rows of Table~\ref{tab:injrec}. In the absence of contamination the framework is unbiased (the residual $\pm$7--9\,ppm ``biases'' are the projection of the single fixed noise realization onto each template, $\approx$0.3$\sigma_A$, and are constant across the injection grid), has unit throughput, nominal coverage, and zero false positives, and GPTLS and the offset-only model are identical to within evidence noise, so the GP neither absorbs signal nor widens the posteriors when there is nothing to model. The comparison with the real four \Tc\ epochs quantifies the cost of TRAPPIST-1's contamination after GP marginalization. The posterior half-width $\sigma_A$ is 25\,ppm on white noise against 41\,ppm on the real data for \co\ (25 versus 33\,ppm for \water, 28 versus 47\,ppm for CH$_4$), and the $\dlnz > 5$ \co\ threshold is 96\,ppm against 171\,ppm, a factor of 1.8. For \water\ and CH$_4$ the real thresholds are offset-assisted and not comparable. The white-noise values, 94 and 102\,ppm, are the honest four-transit sensitivities for those templates on these error bars, on which the three templates are comparably sensitive (within 10 per cent). On the posterior criterion the white-noise thresholds are 79, 65, and 77\,ppm for $z > 3$ (\co, \water, CH$_4$), at or slightly below the $\dlnz > 3$ thresholds of the same rows, as on the real data (135 against 146\,ppm for \co). The lower CH$_4$ sensitivity seen on the \Tb\ archive (Table~\ref{tab:wn_scaling}) is intrinsic to the template on those error bars, not a contamination effect.

\begin{fixedtable}{White-noise scaling study, giving the $\dlnz > 5$ detection thresholds (ppm) on white-noise realizations of the \Tb\ archive with the same subsets, error bars, and configuration as Table~\ref{tab:scaling}, and the ratio of the real-data threshold to the white-noise one. \water\ ratios are below unity because the real thresholds are floor-assisted (Section~\ref{sec:injrec_h2o}). CH$_4$ ratios are given where both thresholds are within the grid.}{tab:wn_scaling}
\footnotesize
\setlength{\tabcolsep}{4pt}
\begin{tabular}{cccccccc}
\toprule
 & \multicolumn{2}{c}{\co} & \multicolumn{2}{c}{\water} & \multicolumn{3}{c}{CH$_4$} \\
\cmidrule(lr){2-3}\cmidrule(lr){4-5}\cmidrule(lr){6-8}
$k$ & WN & real/WN & WN & real/WN & WN & real & real/WN \\
\midrule
2  & 172 & --   & 166 & 0.89 & $>$200 & $>$200 & -- \\
3  & 121 & 1.32 & 125 & 0.82 & 181 & $>$200 & -- \\
4  & 115 & 1.18 & 117 & 0.68 & 155 & 189 & 1.22 \\
5  & 94  & 1.24 & 104 & 0.58 & 149 & 173 & 1.16 \\
6  & 98  & 1.09 & 89  & 0.57 & 138 & 156 & 1.13 \\
7  & 79  & 1.19 & 82  & 0.51 & 128 & 153 & 1.19 \\
8  & 76  & 1.20 & 74  & 0.53 & 116 & 140 & 1.21 \\
9  & 72  & 1.10 & 71  & 0.38 & 113 & 134 & 1.19 \\
10 & 67  & 1.17 & 65  & 0.44 & 110 & 127 & 1.15 \\
\bottomrule
\end{tabular}
\end{fixedtable}

The scaling results (Table~\ref{tab:wn_scaling}) show that the contamination changes the level of the thresholds but not their behaviour. False positives are zero at every $k$ for both criteria, all templates, and both models. The mean $\dlnz$ at zero injection drifts from $-1.4$ at $k=2$ to $-2.3$ at $k=10$, identical to the real data, so the Occam floor is a pure prior-volume effect of the evidence statistic. The fitted exponents ($\dlnz > 5$, $k = 4$--10) are $-0.56$ for \co\ and $-0.41$ for CH$_4$ on white noise against $-0.60$ and $-0.43$ on the real data, indistinguishable within the uncertainties, while for \water\ the white-noise exponent of $-0.64$ against the real $-1.15$ confirms that the real-data steepness is the floor becoming detectable. The real \co\ threshold exceeds the white-noise one by a factor 1.1--1.3 at every $k$ (1.2 for a typical four-transit set, against 1.8 for the actual four \Tc\ epochs, two of which are flare-affected, so the price is epoch-dependent). It is paid in posterior width, through marginalization over GP realizations that could mimic the template, and not in the precision of the recovered amplitude. The subset-to-subset scatter of the zero-injection \co\ amplitude is the same on real and white-noise data (30 versus 29\,ppm at $k=4$, 14 versus 12\,ppm at $k=10$, a ratio of 1.05), whereas for \water\ and CH$_4$ it is $\approx$1.5 times larger on the real data, the floor's epoch-to-epoch variance.

Finally, the white-noise CH$_4$ runs show a persistent $-20$\,ppm offset at every $k$. Because the subset-averaged null mean depends on only 17 base epochs (one noise realization each), it carries an irreducible uncertainty of $\approx\sigma_{\rm single}/\sqrt{17} \approx 15$\,ppm, of which $-20$\,ppm is a 1.3$\sigma$ fluctuation. The same finite-pool uncertainty applies to every $\mu_{\rm b}$ in Table~\ref{tab:null}. The CH$_4$ ``floor'' of $-10$\,ppm on the real \Tb\ archive is consistent with zero, whereas the \water\ floor of $+55$\,ppm is not.

\section{The 2--5\,$\mu$m wavelength range}
\label{app:red}

Because the stellar contamination is concentrated shortward of 2\um\ (Section~\ref{sec:variability}), we repeated the main analysis on the 2--5\um\ range alone, with the same templates, priors, sampler settings, injection grids, control sample, and limit procedure. Table~\ref{tab:red} summarizes the results for GPTLS. Three points are worth recording. First, for \co\ the thresholds are about 10\,ppm higher than on the full range. Dropping the blue half removes the most contaminated wavelengths but also part of the band information, and the two effects nearly cancel, so the amplitude limit is correspondingly slightly weaker (111 against 90\,ppm). Second, the water offset on \Tc\ falls from $+72$ to $+11$\,ppm and the null mean on airless \Tb\ from $+55$ to $+42$\,ppm, with a wider scatter. The floor is carried by the stellar water bands at 1.4 and 1.85\um\ and is largely removed with them, which is direct evidence that it is stellar water. Third, the methane offset of Section~\ref{sec:injrec_ch4} disappears. The offset is $+14$\,ppm with full coverage, the null is centred on zero, and \Tc\ shows no excess ($+0.14\sigma$, $p = 0.40$). The offset seen on the full range therefore comes from the methane bands shortward of 2\um.

\begin{fixedtable}{Results of the 2--5\um\ analysis (GPTLS), in the same conventions as Tables~\ref{tab:injrec}, \ref{tab:null}, and \ref{tab:limits} (amplitudes in ppm). The rows give the injection-recovery offset, posterior half-width, coverage, and detection thresholds on the evidence and posterior criteria on the four \Tc\ epochs, then the \Tb\ null mean and scatter, the \Tc\ amplitude, and its offset from the null, and finally the 95 per cent limit on the planetary amplitude with $A_{\rm p} \ge 0$ (floor-marginalized for \co\ and CH$_4$, statistical-only for \water).}{tab:red}
\footnotesize
\setlength{\tabcolsep}{5pt}
\begin{tabular}{lrrr}
\toprule
 & \co & \water & CH$_4$ \\
\midrule
Offset $\beta$                 & $-23$   & $+11$   & $+14$ \\
Half-width $\sigma_A$          & 41      & 42      & 45 \\
Coverage                       & 100\%   & 100\%   & 100\% \\
Threshold $\dlnz > 3$          & 156     & 124     & 131 \\
Threshold $\dlnz > 5$          & 178     & 158     & 156 \\
Threshold $z > 3$              & 146     & 116     & 116 \\
Null mean $\mu_{\rm b}$        & $+4.7$  & $+42.0$ & $+3.8$ \\
Null scatter $\sigma_{\rm b}$  & 43.9    & 56.1    & 57.3 \\
\Tc\ amplitude $A_{\rm c}$     & $-19.8$ & $+10.7$ & $+11.7$ \\
Offset from null $\delta_\sigma$ & $-0.56$ & $-0.56$ & $+0.14$ \\
$p$                            & 0.71    & 0.72    & 0.40 \\
95 per cent limit              & 111     & 89      & 138 \\
\bottomrule
\end{tabular}
\end{fixedtable}

\bsp	
\label{lastpage}
\end{document}